\documentclass{article}

\usepackage{microtype}
\usepackage{graphicx}
\usepackage{subcaption}
\usepackage{booktabs} % for professional tables

\usepackage{hyperref}

\usepackage[preprint]{icml2026}

\usepackage{amsmath}
\usepackage{amssymb}
\usepackage{mathtools}
\usepackage{amsthm}
\usepackage{hyperref}       % hyperlinks
\usepackage{url}            % simple URL typesetting
\usepackage{booktabs}       % professional-quality tables
\usepackage{amsfonts}       % blackboard math symbols
\usepackage{nicefrac}       % compact symbols for 1/2, etc.
\usepackage{microtype}      % microtypography
\usepackage{xcolor}         % colors
\usepackage{enumitem}

\usepackage{graphicx}
\usepackage{caption}
\usepackage{subcaption}
\usepackage{bm}
\usepackage{amsmath,amssymb}
\usepackage{booktabs}
\usepackage{algorithm}
\usepackage[flushleft]{threeparttable}
\usepackage{multirow}
\usepackage{makecell}
\usepackage{makecell}

\usepackage[table]{xcolor}

\usepackage[capitalize,noabbrev]{cleveref}

\theoremstyle{plain}

\theoremstyle{definition}

\theoremstyle{remark}

\definecolor{yellow}{cmyk}{0,0,1,0}

\usepackage[textsize=tiny]{todonotes}

\icmltitlerunning{Open-Sora 2.0}

\begin{document}

\twocolumn[
  \icmltitle{Open-Sora 2.0: Training a Commercial-Level Video Generation Model in \$200k}



  \begin{icmlauthorlist}
    \icmlauthor{Open-Sora Team}{comp}
  \end{icmlauthorlist}

  \icmlaffiliation{comp}{HPC-AI Tech}

  \icmlcorrespondingauthor{Yang You}{youy@comp.nus.edu.sg}

  \icmlkeywords{Video Generation, Diffusion Transformer}

  \vskip 0.3in
]



\printAffiliationsAndNotice{}  % no special notice (required even if empty)

\begin{abstract}
Video generation models have achieved remarkable progress in the past year. The quality of AI video continues to improve, but at the cost of larger model size, increased data quantity, and greater demand for training compute. In this report, we present Open-Sora 2.0, a commercial-level video generation model trained for only \$200k. With this model, we demonstrate that the cost of training a top-performing video generation model is highly controllable. We detail all techniques that contribute to this efficiency breakthrough, including data curation, model architecture, training strategy, and system optimization. According to human evaluation results and VBench scores, 
% Open-Sora 2.0 is comparable to global leading video generation models such as OpenAI's Sora, Runway Gen-3 Alpha, and HunyuanVideo. 
Open-Sora 2.0 is comparable to global leading video generation models including the open-source HunyuanVideo and the closed-source Runway Gen-3 Alpha.
By making Open-Sora 2.0 fully open-source, we aim to democratize access to advanced video generation technology, fostering broader innovation and creativity in content creation. 
\end{abstract}
\section{Introduction}
\label{sec:intro}

\begin{figure}[t]
    \begin{center}
        \centerline{\includegraphics[width=\columnwidth, trim={0cm 0cm 0cm 0cm}, clip]{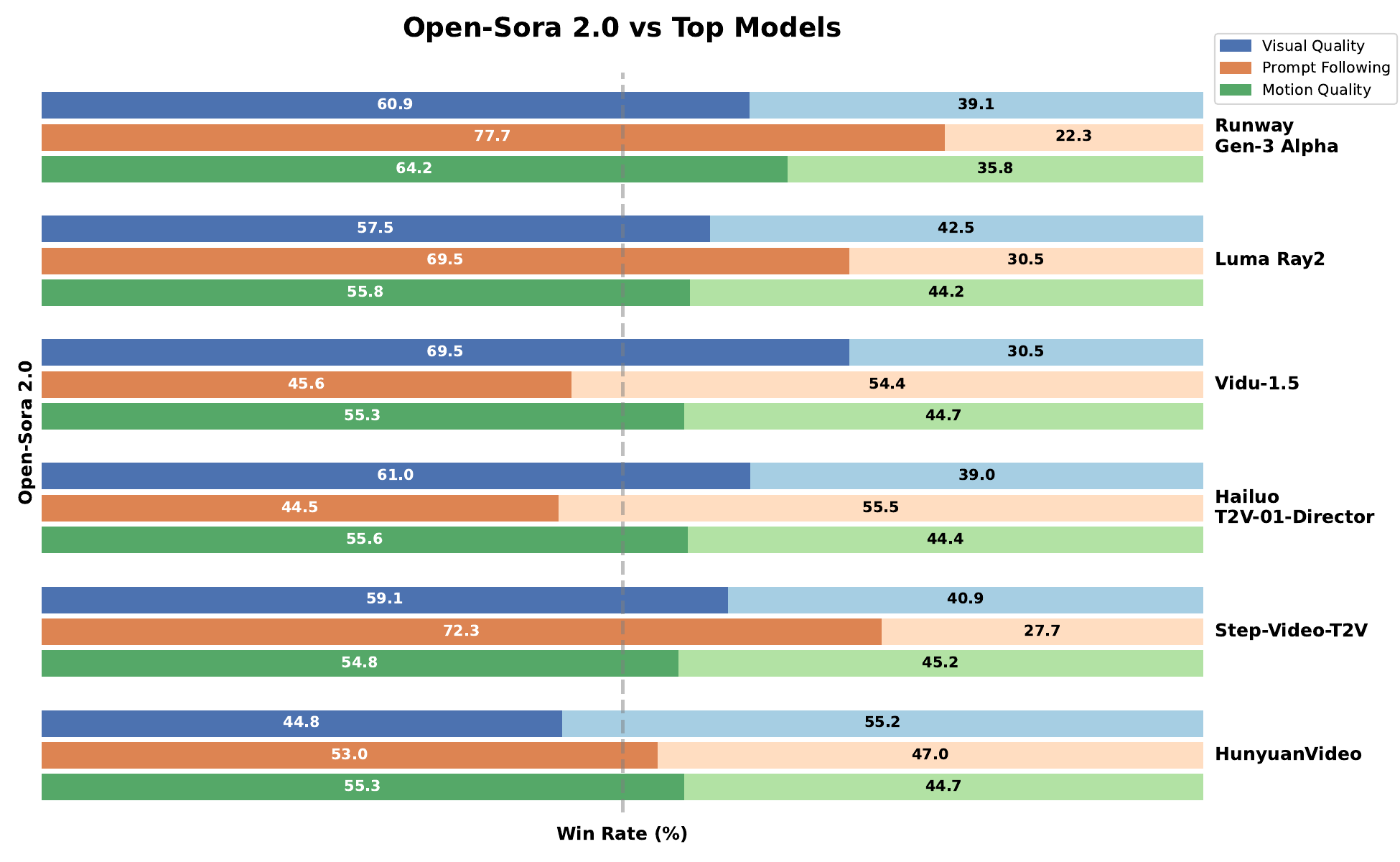}}
        \caption{Human preference evaluation of Open-Sora 2.0 against other leading video generation models. Win rate represents the percentage of comparisons where our model was preferred over the competing model. The evaluation is conducted on 100 prompts carefully designed to cover three key aspects: 1) visual quality, 2) prompt adherence, and 3) motion quality. Results show that our model performs favorably against other top-performing models in all three aspects.}
        \label{fig:winrate}
    \end{center}
    \vskip -30pt
\end{figure}
% Diffusion models have demonstrated huge power in image generation.
The past year has witnessed an explosion of video generation models. Since the emergence of OpenAI's Sora~\cite{videoworldsimulators2024} in February 2024, a series of video generation models---either open-source~\cite{zheng2024open, lin2024open,ma2025step,kong2024hunyuanvideo,hong2022cogvideo} or proprietary~\cite{polyak2024moviegen,runwayml_gen3,luma_dream_machine}---have appeared at an unprecedented pace, each striving to achieve "Sora-level" generation quality. While the quality of generated videos continues to improve, there is a clear trend toward rapid growth in model size, data quantity, and computing resources. Following the success of scaling large language models (LLMs)~\cite{hoffmann2022training,kaplan2020scaling}, researchers are now applying similar scaling principles~\cite{kong2024hunyuanvideo} to video generation, converging on similar model architectures and training techniques.

In this paper, however, we show that a top-performing video generation model can be trained at a highly controlled cost, offering new insights into cost-effective training and efficiency optimization. We build Open-Sora 2.0, a commercial-level video generation model trained for only \$200k. As shown in Figure~\ref{fig:cost-comp}, the training cost is 5-10 times lower than comparable models like MovieGen~\cite{polyak2024moviegen} and Step-Video-T2V~\cite{ma2025step} based on a fair comparison. This remarkable cost efficiency stems from our joint optimization of data curation, training strategy, and AI infrastructure---all of which are detailed in the following sections.

We show the human preference evaluation of Open-Sora 2.0 and other global leading video generation models in Figure~\ref{fig:winrate}. The models for comparison include proprietary models like Runway Gen-3 Alpha~\cite{runwayml_gen3}, Luma Ray2~\cite{luma_dream_machine} and open-source models like HunyuanVideo~\cite{kong2024hunyuanvideo} and Step-Video-T2V~\cite{ma2025step}. We evaluate these models in three important aspects: 1) visual quality, 2) prompt adherence, and 3) motion quality. Despite the low cost of our model, it outperforms these top-performing models in at least two aspects out of the three, demonstrating its strong potential for commercial deployment.

% This paper is structured as follows. In Section~\ref{sec:data}, we elaborate on our data strategy, including our hierarchical data filtering system and annotation methods. Section~\ref{sec:model} details our model architecture, covering both our novel Video DC-AE autoencoder design and the DiT architecture. Section~\ref{sec:train} explores our cost-effective training methodology, which enables commercial-level quality at just \$200k. Section~\ref{sec:condition} presents our conditioning approaches, including image-to-video and motion control techniques.
% % , and inference time scaling.
% Section~\ref{sec:system} outlines the system optimizations that maximize training efficiency, and Section~\ref{sec:performance} evaluates our model's performance against leading video generation models. Finally, we conclude with a summary of our contributions and insights for future research.
\section{Data}
\label{sec:data}
Our goal for data is to build a hierarchical data pyramid, catering to the requirement of the progressive training process. To this end, we develop a collection of filters that function distinctly from each other, aiming to tackle various types of data detections. By progressively strengthening the degree of filtration, we can obtain subsets of smaller sizes but higher purity and quality. For completeness, we further conduct a statistical analysis of some key attributes of the collected video data, including distributions of visual and textual attributes in Appendix~\ref{sec:data_stat}.

% In Section~\ref{sec:data_filtering}, we first elaborate the data filtering system that consists of two main components: preprocessing and score filtering. Then, we introduce data annotation methods in Section~\ref{sec:data_ann}. Finally, we present detailed statistics of the whole dataset in Section~\ref{sec:data_stat}.

\subsection{Data Filtering}
\label{sec:data_filtering}

\begin{figure}[h]
    \begin{center}
        \centerline{
            \includegraphics[width=\columnwidth, trim={2cm 2cm 8.5cm 3.5cm}, clip]{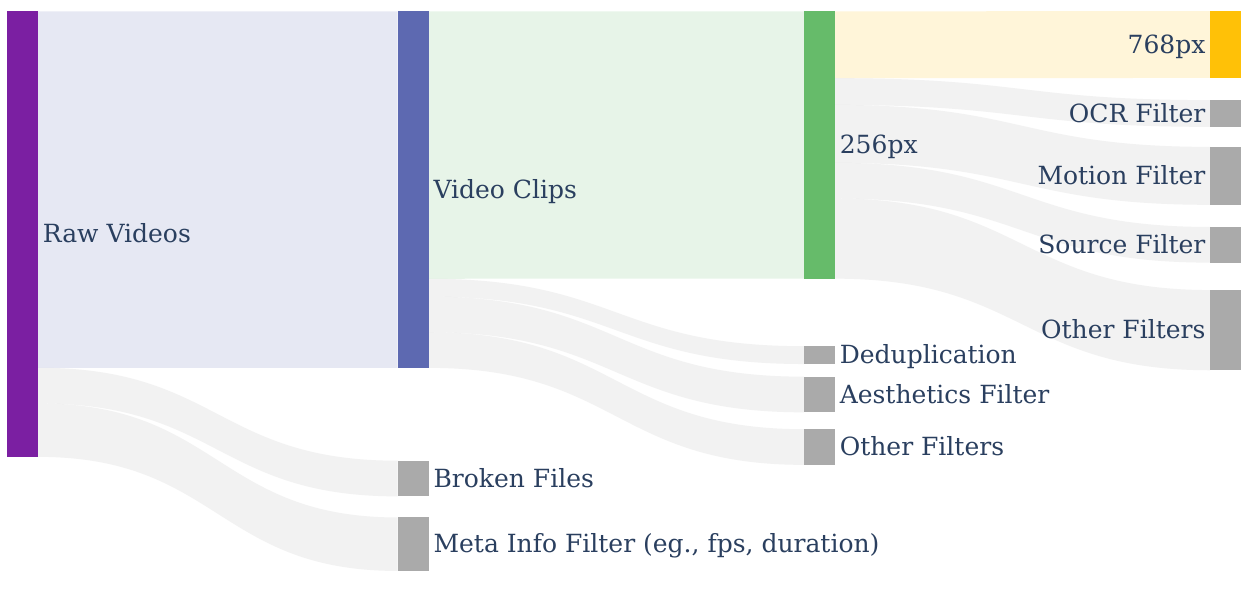}
        }
                \caption{The hierarchical data filtering pipeline. The raw videos are first transformed into trainable video clips. Then, we apply various complimentary score filters to obtain data subsets for each training stage. }
        \label{fig:data_filtering}
    \end{center}
\vskip -20pt
\end{figure}

The hierarchical data filtering system is illustrated in Figure~\ref{fig:data_filtering}. We begin by preprocessing raw videos into video clips, then progressively apply a series of filters, ranging from loose to strict, to construct a structured data pyramid.

\subsubsection{Preprocessing}
The preprocessing phase converts raw videos into short clips suitable for training. During this phase, we first eliminate broken files and raw videos with outlying attributes. Specifically, we filter out videos with a duration of less than 2 seconds, a bit per pixel (bpp) below 0.02, a frame rate (fps) under 16, and an aspect ratio outside the range [1/3, 3], and those with a "Constrained Baseline" profile. Then, we segment raw videos into short continuous clips based on scene scores calculated by libavfilter from FFmpeg~\cite{FFmpeg}. Finally, we process the obtained continuous clips, ensuring that the output clips adhere to specific format constraints: frame rate (fps) below 30, longer dimension not exceeding 1080 pixels, and H.264 codec. Additionally, black borders are removed by clipping the shots. Last, we divide shots exceeding 8 seconds into multiple 8-second clips, while shots shorter than 2 seconds are discarded.

\subsubsection{Score Filtering}
To address the various defects in raw data, we develop a bag of complimentary filters, including 1) aesthetic score; 2) motion score; 3) blur detection; 4) OCR; 5) camera jitter detection, each filter targets a specific aspect of data quality.
% filter functions in its unique aspect and they work jointly as a comprehensive and strong purifying system.
These filters work together as a comprehensive and robust purification system.
Typically, each filter evaluates a sample by assigning a score based on its respective criteria, and the filtering intensity is controlled by a threshold. We introduce all the score-based filters in Appendix~\ref{sec:data_filters}.

\subsection{Data Annotation}
\label{sec:data_ann}
For captioning, we utilize the open-source vision-language model LLaVA-Video~\cite{zhang2024video} to annotate 256px videos. We prompt the model to focus on six aspects for a detailed and comprehensive caption, which are 1) main subjects; 2) subjects' actions; 3) background and environment; 4) lighting condition and atmosphere; 5) camera movement; 6) video style, such as realistic, cinematic, 3D, animation, and so on. For high-resolution 768px training data, we leverage a stronger proprietary model Qwen 2.5 Max~\cite{qwen25} to generate more accurate and semantically aligned captions. We find that Qwen 2.5 Max produces fewer hallucinations and delivers better semantic consistency than LLaVA-Video. For both training and inference, we append the motion score after the caption (detailed in Section~\ref{sec:motion_score}).
\section{Model Architecture}
\label{sec:model}

The two key components of a video generation model are the autoencoder and the diffusion transformer. For the autoencoder, our model is initially trained on HunyuanVideo’s VAE and later adapted to our Video DC-AE, whose structure is detailed in Sec.~\ref{sec:autoencoder}. The architecture of the diffusion transformer is presented in Sec.~\ref{sec:arch}.

\subsection{3D Autoencoder}
\label{sec:autoencoder}

\begin{figure}[t]
    \begin{center}
        \centerline{\includegraphics[width=0.75\columnwidth]{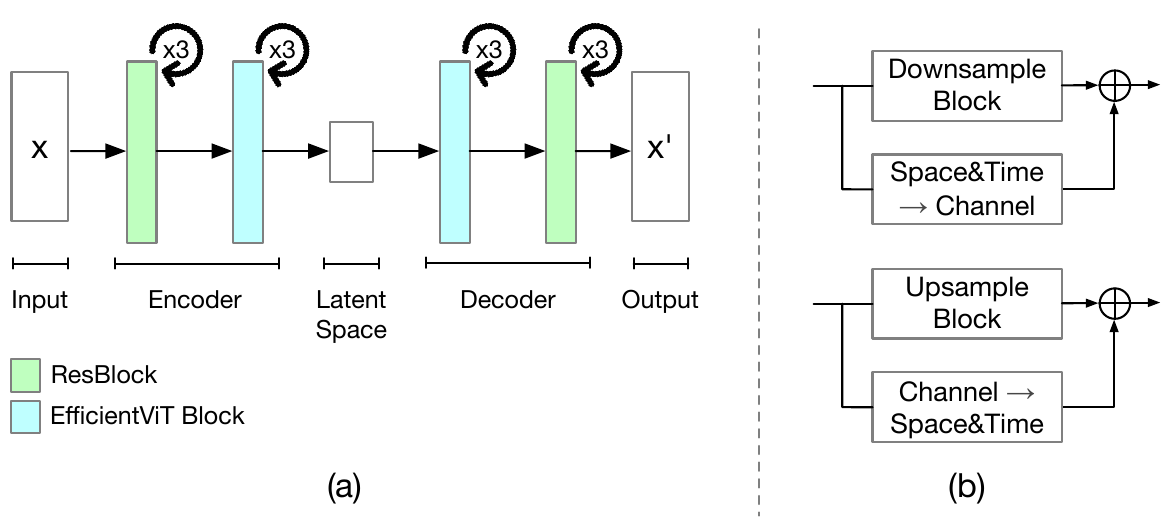}}
        \caption{Architecture of Video DC-AE. (a) Overview of Video DC-AE: Each block in encoder introduces spatial downsampling, while temporal downsampling occurs at blocks 4 and 5, with a corresponding symmetric structure in the decoder. (b) Residual Connection in Video DC-AE Blocks.}
        \label{fig:arch_ae}
    \end{center}
    \vskip -30pt
\end{figure}

We begin by leveraging the open-source HunyuanVideo VAE~\cite{kong2024hunyuanvideo} as the initial autoencoder. To further reduce both training and inference costs, we develop a video autoencoder with deep compression (Video DC-AE, named after \cite{chen2024deep}) that improves efficiency while maintaining high reconstruction fidelity.

\begin{figure}[t]
    \begin{center}
        \centerline{\includegraphics[width=\columnwidth]{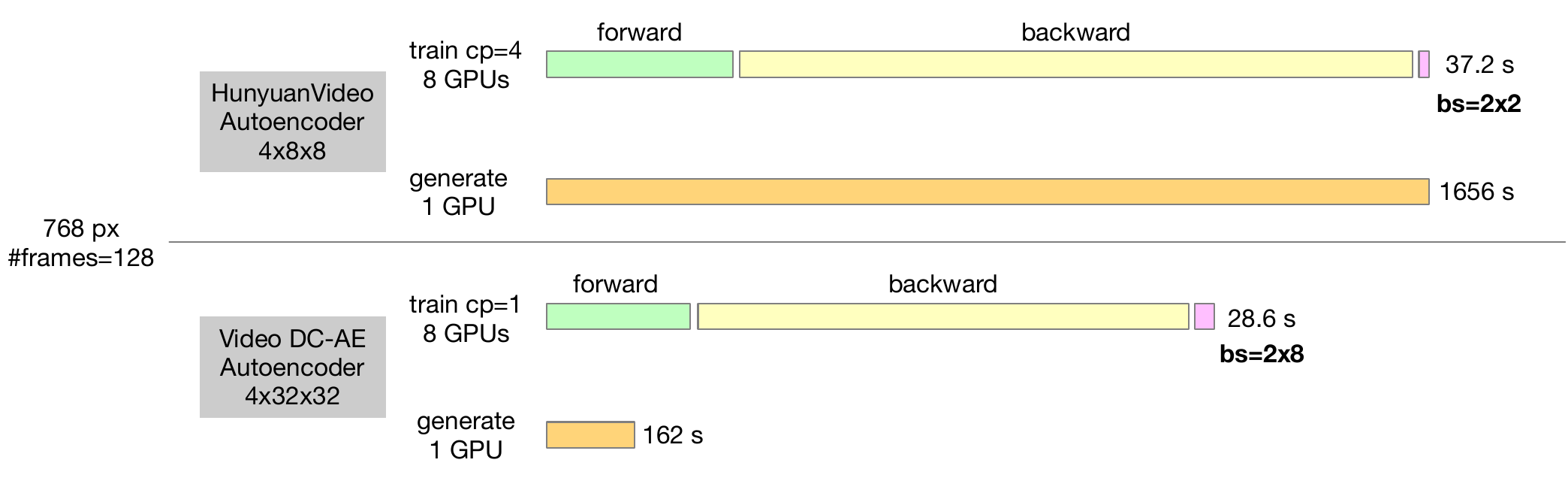}}
        \caption{Training and Inference Speed Comparison at 768px. This figure compares the training and inference speeds at 768px resolution using HunyuanVideo VAE ($4\times8\times8$, patch size 2) and Video DC-AE ($4\times32\times32$, patch size 1). The results demonstrate that Video DC-AE achieves significantly higher efficiency in both training and inference (50 steps) compared to HunyuanVideo VAE, highlighting its advantage in high-resolution video generation.}
        \label{fig:hcae-speed}
    \end{center}
    \vskip -30pt
\end{figure}

HunyuanVideo VAE achieves a compression ratio of  $4 \times 8 \times 8$. Under that compression ratio, even with a patch size of $2 \times2$, our generation model still needs to process around 115K tokens per training video, resulting in considerable computational demands. To address the above challenge, we identify potential redundancy in the spatial dimension and propose increasing the spatial compression ratio to 32 while keeping the temporal compression ratio fixed at 4. As shown in Figure~\ref{fig:hcae-speed}, during training, our Video DC-AE with a $4 \times 32 \times 32$ compression ratio and a patch size of $1 \times 1 \times 1$ enables training on 32-frame, 256px videos with a latent shape of $8 \times 8 \times 8$. This configuration preserves a balanced level of information across all dimensions and yields a $5.2\times$ improvement in training throughput. For generation, for instance, compared to HunyuanVideo VAE, our Video DC-AE reduces the token count from 76K to 19K for a 5-second, 24fps, 768px video. A detailed derivation of the token count reduction is provided in Appendix~\ref{sec:token_count_reduction}. In addition, inference speed improves by over $10\times$, making this approach highly efficient. 

\citet{chen2024deep} introduce DC-AE as an effective approach to achieve higher downsampling ratios while maintaining reconstruction quality. While the original DC-AE is primarily optimized for image encoding, we extend its architecture to support video encoding by incorporating temporal compression mechanisms.

As shown in Figure~\ref{fig:arch_ae}(a), our Video DC-AE encoder consists of three residual blocks followed by three EfficientViT blocks~\cite{cai2023efficientvit}, with the decoder adopting a symmetrical structure. 
The first five encoder blocks and the last five decoder blocks function as downsampling and upsampling layers, respectively.

To adapt DC-AE for video encoding, we replace 2D operations (e.g., convolutions, normalization) with 3D operations.
Additionally, we introduce temporal compression in the last two downsampling blocks (blocks 4 and 5) of the encoder, and temporal upsampling in the first two upsampling blocks (blocks 2 and 3) of the decoder to reconstruct temporal information effectively. Moreover, we follow DC-AE to introduce special residual blocks to connect the downsample and upsample blocks. This is because \citet{chen2024deep} identifies gradient propagation issues in these blocks, such that without the residuals, it is especially difficult to train high-compression autoencoders.

As shown in Figure~\ref{fig:arch_ae}(b), our downsample residual blocks follow DC-AE’s pixel-shuffling strategy, redistributing pixels from the spatial and temporal dimensions into the channel dimension, followed by channel-wise averaging to achieve the desired compression. 
The upsample residual blocks perform the inverse operation by first duplicating channels, then redistributing them back into the spatial and temporal dimensions to reconstruct the original structure.

% We train Video DC-AE from scratch and evaluate its reconstruction quality, with results summarized in Table~\ref{tab:vae-perf}. We compare our model with OpenSora 1.2 VAE, the recently open-source StepVideo VAE\cite{ma2025step}, and HunyuanVideo VAE\cite{kong2024hunyuanvideo}. The results indicate that Video DC-AE achieves competitive performance, with only slight degradation in LPIPS~\cite{zhang2018unreasonable} scores compared to the top-performing HunyuanVideo VAE, while maintaining strong PSNR and SSIM scores. Furthermore, although the 256-channel version of Video DC-AE achieves superior reconstruction quality, we select the 128-channel variant to facilitate faster adaptation in the video generation model due to its smaller channel size (see Section~\ref{sec:hc_ae} for further details).

\begin{table}[t]
\centering
\caption{\textbf{Auto-encoder reconstruction performance comparison}. The Video DC-AE highlighted with yellow background is selected for generative model adaptation.}
\resizebox{\columnwidth}{!}{%
\begin{tabular}{@{}llcccccc@{}}
\toprule
Model & Down. (TxHxW) & Info. Down. & Channel & Causal & \multicolumn{1}{l}{LPIPS↓} & \multicolumn{1}{l}{PSNR↑} & \multicolumn{1}{l}{SSIM↑} \\ \midrule
Open-Sora 1.2~\cite{zheng2024open} & $4\times 8\times 8$ & 192 & 4   & $\checkmark$ & 0.161 & 27.504 & 0.756 \\
StepVideo VAE~\cite{ma2025step}  & $8\times 16\times 16$ & 96 & 64  & $\checkmark$ & 0.082 & 28.719 & 0.818 \\
HunyuanVideo VAE~\cite{kong2024hunyuanvideo}    & $4\times 8\times 8$ & 48 & 16  & $\checkmark$ & \textbf{0.046} & 30.240 & 0.856 \\ \midrule
\rowcolor{yellow} Video DC-AE    & $4\times 32\times 32$ & 96 & 128 & $\times$ & 0.051 & 30.538 & 0.863 \\
Video DC-AE     & $4\times 32\times 32$ & 48 & 256 & $\times$ & 0.049 & \textbf{30.777} & \textbf{0.872} \\ \bottomrule
\end{tabular}%
}
\vskip -20pt
\label{tab:vae-perf}
\end{table}

We train Video DC-AE from scratch and evaluate its reconstruction quality, with results summarized in Table~\ref{tab:vae-perf}. Our model is compared against Open-Sora 1.2 VAE, StepVideo VAE~\cite{ma2025step}, and HunyuanVideo VAE~\cite{kong2024hunyuanvideo}. The results demonstrate that Video DC-AE achieves competitive performance, with only minor degradation in LPIPS~\cite{zhang2018unreasonable} scores compared to the best-performing HunyuanVideo VAE, while maintaining strong PSNR and SSIM scores.

Furthermore, while the 256-channel version of Video DC-AE offers higher reconstruction quality, we opt for the 128-channel variant in the video generation model to enable faster adaptation due to its reduced channel size (see Section~\ref{sec:hc_ae} for further details).

\subsection{DiT Architecture}
\label{sec:arch}

\begin{figure}[ht]
  % \vskip -15pt
  \begin{center}
    \centerline{\includegraphics[width=0.5\columnwidth, trim={0cm 0.5cm 0cm 0.5cm}, clip]{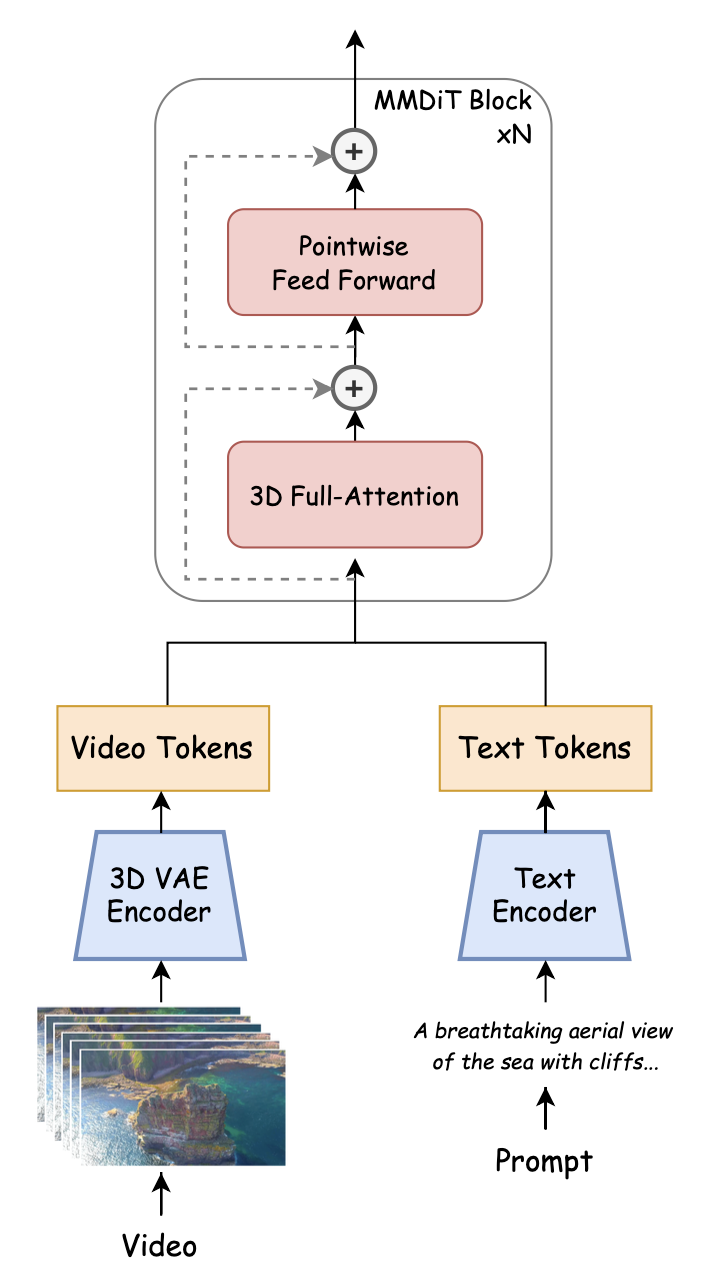}}
    \caption{
      Open-Sora diffusion transformer architecture.
    }
    \label{fig:arch}
  \end{center}
\vskip -30pt
\end{figure}

To achieve higher video quality, we employ full attention to capture long-range dependencies effectively. After encoding the input through the autoencoder, we patchify the latent representations to enhance computational efficiency and improve model learning. 
Sana~\cite{xie2024sana} suggests that a patch size of 1 yields better training stability and finer details in video generation.
With Video DC-AE's high compression ratio, we can afford to reduce the patch size to 1 (i.e., no patch at all), whereas a patch size of 2 is still required when the HunyuanVideo autoencoder is used in our generation model.
% Specifically, we use a patch size of 2 for the HunyuanVideo autoencoder and 1 for the Video DC-AE, following prior research suggesting that a patch size of 1 yields better training stability and finer details in video generation.

Inspired by FLUX’s MMDiT~\cite{flux2024}, we employ a hybrid transformer architecture that incorporates both dual-stream and single-stream processing blocks. In the dual-stream blocks, text and video information are processed separately to facilitate more effective feature extraction within each modality. Subsequently, single-stream blocks integrate these features to facilitate effective cross-modal interactions. To further enhance the model’s ability to capture spatial and temporal information, we apply 3D RoPE (Rotary Position Embedding)~\cite{su2024roformer}, which extends traditional positional encoding to three-dimensional space, allowing the model to better represent motion dynamics across time. For text encoding, we leverage T5-XXL~\cite{chung2024scaling} and CLIP-Large~\cite{Radford2021LearningTV}, two high-capacity pretrained models known for their strong semantic understanding. T5-XXL captures complex textual semantics, while CLIP-Large improves alignment between text and visual concepts, leading to more accurate prompt adherence in video generation.

An overview of the model architecture is illustrated in Figure~\ref{fig:arch}, and the detailed architectural specifications are presented in Table~\ref{tab:model-arch} in Appendix~\ref{sec:model_arch_hyperparameter}.

\section{Model Training}
\label{sec:train}

A commercial-grade video generation model typically requires more than 10 billion parameters and O(100M) training samples. To mitigate the substantial computational cost associated with training, we propose a cost-effective training pipeline, as outlined in Table~\ref{tab:training-stage}. Our approach consists of three key stages: (1) training a text-to-video (T2V) model on low-resolution video data, (2) training an image-to-video (I2V) model on low-resolution video data, and (3) fine-tuning an I2V model on high-resolution videos.

\begin{table}[t]
\centering
\caption{Training Configurations and Cost Breakdown. This table presents the training configurations across different stages and the total cost for a single full training run, assuming the rental price of H200 is \$2 per GPU hour.}
\label{tab:training-stage}
\vskip -10pt
  \begin{center}
    \begin{scriptsize}
      \begin{sc}
        \resizebox{\columnwidth}{!}{%
        \begin{tabular}{lcccccr}
        \toprule
        \makecell{Training \\ Stage} & Dataset & CP & \#iters & \#GPUs & \#GPU day & USD  \\ \midrule
        256px T2V      & 70M     & 1  & 85k     & 224    & 2240 & \$107.5k  \\
        256px T/I2V    & 10M     & 1  & 13k     & 192    & 384  & \$18.4k   \\
        768px T/I2V    & 5M      & 4  & 13k     & 192    & 1536 & \$73.7k    \\ \midrule
        Total          &         &    &         &        & 4160 & \$199.6k  \\\bottomrule
        \end{tabular}%
        }
      \end{sc}
    \end{scriptsize}
  \end{center}
  \vskip -20pt
\end{table}

\subsection{Efficient Training Strategy}

Our budget-conscious training strategy emphasizes the following four key aspects:

\subsubsection{Leveraging Open-Source Image Models} Prior research~\cite{ma2024latte,hong2022cogvideo,zheng2024open} has demonstrated that pretraining on image datasets can significantly accelerate video model training. To avoid the high cost of training an image model from scratch, we leverage Flux~\cite{flux2024}, an open-source state-of-the-art text-to-image model with 11 billion parameters, which provides sufficient capacity to generate high-quality videos. We initialize our T2V model using Flux and empirically find that, despite it being a distilled model, this initialization is effective for further training.

\subsubsection{High-Quality Training Data} Inspired by the efficient image training strategies from PixArt~\cite{chen2023pixart}, we hypothesize that high-quality video data can substantially enhance training efficiency. Consequently, we curate a high-quality subset from a large-scale dataset for low-resolution training. For high-resolution fine-tuning, we impose stricter selection criteria to ensure superior video quality.

\subsubsection{Learning Motion in Low-Resolution} Training a commercial video generation model is computationally expensive, particularly at high resolutions. To mitigate this cost, we first train on 256px resolution videos, allowing the model to learn diverse motion patterns efficiently. However, we observe that while the model captures motion effectively, low-resolution outputs tend to be blurry. Increasing the resolution significantly improves perceptual quality and human evaluative feedback.

As shown in Table~\ref{tab:stage1-bs} and Table~\ref{tab:stage2-bs} in Appendix~\ref{sec:multi-bucket-training}, training on a 129-frame video at 768px resolution is 40 times slower than at 256px. This performance gap arises due to the quadratic computational complexity of self-attention mechanisms as the number of tokens increases. To optimize efficiency, we allocate the majority of computational resources to low-resolution training, reducing the need for expensive high-resolution computations.

\subsubsection{Image-to-Video Models Facilitate Resolution Adaptation} We find that adapting a model from 256px to 768px resolution is significantly more efficient using an I2V approach compared to T2V. We hypothesize that conditioning the model on a static image allows it to focus more on motion generation, a capability that is well-learned during low-resolution training.

Based on this observation, we prioritize training an I2V model at high resolution. During inference, we first generate an image from a text prompt and subsequently synthesize a video conditioned on both the image and the text. During training, T/I2V training at low resolution followed by brief fine-tuning on high-resolution videos yields high-quality results with minimal additional training.

\begin{figure}[h]
    \begin{center}
        \centerline{\includegraphics[width=\columnwidth]{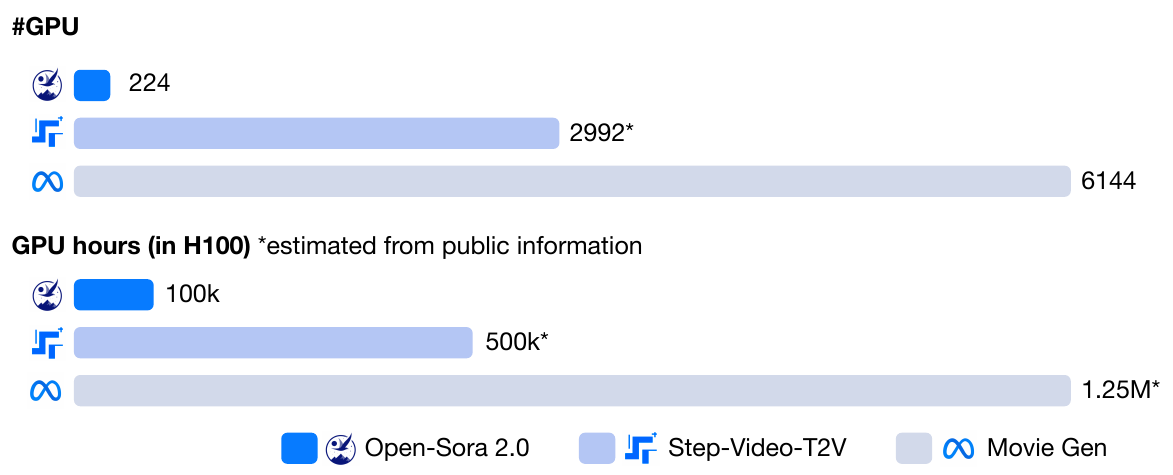}}
        \caption{Training Cost Comparison of Different Video Generation Models. Values marked with~* are estimated based on publicly available information. Our Open-Sora 2.0 is trained at 5–10× lower training cost.}
        \label{fig:cost-comp}
    \end{center}
    \vskip -30pt
\end{figure}

With our proposed approach, we successfully constrain the one-time training cost to \$200K, as detailed in Table~\ref{tab:training-stage}. We estimate the training costs of Movie Gen~\cite{polyak2024moviegen} and Step-Video-T2V~\cite{ma2025step} based on publicly available information, as summarized in Figure~\ref{fig:cost-comp}. Our model achieves 5–10× lower training costs. We hope that these insights will contribute to reducing the training cost of high-quality video generation models in future research.
% \begin{table}[h]
% \centering
% \resizebox{0.8\textwidth}{!}{%
% \begin{tabular}{@{}llll@{}}
% \toprule
% Model          & \#GPUs & GPU Hours & Cost (Single Run) \\ \midrule
% Movie Gen \cite{polyak2024moviegen}      & 6144   & 1.25M*    & \$2.5M*              \\
% Step-Video-T2V \cite{ma2025step} & 2992*  & 500k*     & \$1M*                \\
% Open Sora 2.0  & 224    & 100k      & \$200k              \\ \bottomrule
% \end{tabular}%
% }
% \vspace{5pt}
% \caption{\textbf{Training Cost Comparison of Different Video Generation Models}. Values marked with * are estimated based on publicly available information. Our Open-Sora 2.0 is trained at 5–10× lower training cost.}
% \label{tab:cost-comp}
% \end{table}
% \begin{table}[h]
% \centering
% \resizebox{0.8\textwidth}{!}{%
% \begin{tabular}{@{}lll@{}}
% \toprule
% Model          & \#GPUs & GPU Hours \\ \midrule
% Movie Gen \cite{polyak2024moviegen}      & 6144   & 1.25M*                  \\
% Step-Video-T2V \cite{ma2025step} & 2992*  & 500k*                  \\
% Open Sora 2.0  & 224    & 100k                 \\ \bottomrule
% \end{tabular}%
% }
% \vspace{5pt}
% \caption{\textbf{Training Cost Comparison of Different Video Generation Models}. Values marked with * are estimated based on publicly available information. Our Open-Sora 2.0 is trained at 5–10× lower training cost.}
% \label{tab:cost-comp}
% \end{table}

\subsection{Training Setting}

Our training setup is largely based on Open-Sora 1.2 \cite{zheng2024open}. We adopt flow matching \cite{lipman2022flow} as our primary training objective and utilize the AdamW \cite{loshchilov2017decoupled} optimizer with $\beta$ values of (0.9, 0.999) and an $\epsilon$ value of $1\times 10^{-15}$. No weight decay is applied. The learning rate is set to $5\times 10^{-5}$ for the first 40k steps of Stages 1 and 2, followed by a decay to $3\times 10^{-5}$ for the final 45k steps. For Stage 3, it is reduced to $1\times 10^{-5}$. Gradient norm clipping is applied with a threshold of 1.

\subsubsection{Training Objective}

Our flow matching approach is similar to that used in Stable Diffusion 3 \cite{esser2024scaling}. 
We denote the video latent as $X_0$, and Gaussian noise $X_1\sim\mathcal{N}(0,1)$. The model $f_\theta$ takes as input an interpolated latent $X_t=(1-t)X_0+tX_1$ and is trained to predict the velocity component $X_0-X_1$. The corresponding loss function is formulated as:

$$\mathcal{L}=\mathbb{E}_{t,X_0,X_1}[\|f_\theta(X_t,t,y)-(X_0-X_1)\|],$$

where $y$ represents the conditioning input (text and/or image). The timestep $t$ is first sampled from a logit-normal distribution and then scaled according to the shape of $X_0$. Given that higher-resolution and longer-duration videos are more susceptible to noise, we apply the following transformation: $t\leftarrow \frac{\alpha t}{1 + (\alpha -1)t}$ during training and inference, where $\alpha$ is proportional to the product $T\times H\times W$.

\subsubsection{Multi-Bucket Training}
Following the methodology in Open-Sora 1.2~\cite{zheng2024open}, we adopt multi-bucket training to efficiently handle videos of varying frame counts, resolutions, and aspect ratios within the same batch. This strategy optimizes GPU utilization by dynamically assigning batch sizes based on video characteristics. We conduct a search on H200 GPUs to determine batch sizes used for different training configurations. The specific batch sizes and hyperparameter searching methods are detailed in Appendix~\ref{sec:multi-bucket-training}. 

\subsection{High Compression Autoencoder Adaptation}
\label{sec:hc_ae}

As discussed earlier, the high computational cost of training video generation models arises from the large number of tokens and the dominance of attention computation. Generating a one-minute video in a single pass would incur an extremely high cost in the future. To further reduce training expenses, we explore training video generation models with high-compression autoencoders (Video DC-AEs).

\subsubsection{Challenges in Training Video Autoencoders.} With the growing interest in video generation models, we observe that despite variations in architecture, downsampling ratios, number of channels, and training losses, video autoencoders trained with similar information downsampling ratios exhibit comparable performance (evaluated at 256px resolution) \cite{chen2024deep}. 
We define the information downsampling ratio, $D_{\text{info}}$ as follows:
\[ 
    D_{\text{info}} = \frac{D_{\text{T}} \times D_{\text{H}} \times D_{\text{W}} \times C_{\text{in}}}{C_{\text{out}}}
\]
where $C_{\text{in}}$ and $C_{\text{out}}$ are the number of input and output channels (note that $C_{\text{in}} = 3$ for videos using RGB channels). 
We do not consider the actual storage sizes due to possible confounders of different data types (such as float32 or bfloat16) used.

We hypothesize that such a link between $D_{\text{info}}$ and the reconstruction performance indicates a fundamental information compression lower bound for a given resolution, thereby constraining the achievable compression ratio unless the number of channels is adjusted. Our two video autoencoders of 256 channels and 128 channels are designed to match the $D_{\text{info}}$ of the HunyuanVideo VAE and the StepVideo VAE respectively.

However, training a deep video compression network is challenging. Prior work~\cite{chen2024deep} identifies gradient propagation issues in downsample and upsample blocks. To address this, they introduce residual connections in these blocks using pixel shuffle and unshuffle operations. We verify that training the HunyuanVideo VAE architecture with residual connections improves performance. Building on this, we incorporate temporal downsampling into DC-AE to construct Video DC-AE.

Additionally, we find that maintaining similar reconstruction performance requires a 4$\times$ increase in channels when doubling the height and width compression ratio, whereas adding a 4$\times$ temporal compression on top of the original image DC-AE (no temporal compression) incurs no additional channels possibly due to redundancy in temporal information. Based on these insights, we train a Video DC-AE with a 2x $D_{\text{info}}$ and 16x $D_{\text{token}}$ as compared to the HunyuanVideo VAE while maintaining comparable performances.

\subsubsection{Challenges in Training the Generation Model.}
While training throughput for the diffusion model improves, we find that high-compression autoencoders—especially those with larger latent channel dimensions—slow down convergence in training the diffusion model. Prior studies~\cite{yao2025reconstruction, xie2024sana} show that when training image diffusion models on AEs with better reconstruction but higher latent dimensionality, the generation quality often degrades. Additionally, Theorem A.7 in \cite{chen2025masked} suggests that, under strong assumptions, increasing the channel size by $k$ requires $k^5$ more data to achieve comparable performance. Meanwhile, \citet{yao2025reconstruction} demonstrates that introducing a distillation loss between a pretrained image foundation model and the VAE latents can significantly accelerate diffusion training.

Based on these findings, we hypothesize that current VAE training frameworks struggle to optimize latent space structures for video generation when channel sizes increase. 
Although reconstruction ability sets an upper bound on generation quality, a well-structured latent space is more critical for effective video synthesis.

\subsubsection{Strategies for Using High-Compression Autoencoders.}

Despite the challenges of training with high-compression autoencoders, we adopt the following strategies to construct a fast video generation model:

\begin{enumerate}
    \item \textbf{Latent Space Distillation}: After training Video DC-AE, we apply a distillation loss to align the third-layer latents with DINOv2~\cite{oquab2023dinov2}, leveraging their similar latent shapes.
    \item \textbf{Efficient Adaptation for Diffusion Models}: Following PixArt, diffusion models can be adapted to different autoencoders. We reinitialize the input and output layers of the model for adaptation. We find that adaptation to high-compression autoencoders behaves a bit differently. Semantic structures adapt quickly, but video outputs appear blurry. 
    \item \textbf{Prioritizing I2V Training}: We observe that I2V models adapt more efficiently than T2V models when switching to high-compression AEs. Thus, we focus on training an I2V model for this setting.
    \item \textbf{Tiling in video encoding}: High compression AEs trained at low resolutions experience performance degradation when reconstructing high-resolution videos \cite{chen2024deep}. Although we could have fine-tuned the Video DC-AE at high resolutions, we re-use the tiling code by \citet{kong2024hunyuanvideo} to save training resources. 
\end{enumerate}

We first train the model on short videos (up to 33 frames) for 17K iterations using 20M samples across 160 GPUs. Then, we extend training to long videos (up to 128 frames) for 8K iterations on 2M samples. The final model achieves a loss of 0.5, compared to 0.1 for the untrained model. However, due to computational constraints, training does not fully converge. As shown in Figure~\ref{fig:ae-comp} in Appendix~\ref{appendix:video_gen_ae}, while the fast video generation model underperforms the original, it still captures spatial-temporal relationships. We release this model to the research community for further exploration.

\subsubsection{Discussion and Future Work.} We believe high-compression autoencoders are critical for end-to-end video generation, especially for high-resolution videos that contain substantial redundancy. To further reduce training costs, it is essential to: (1) Optimize autoencoder training to yield better latent spaces for diffusion-based learning. (2) Design high-throughput autoencoders, as autoencoder encoding time remains a major computational bottleneck in generative models using high-compression autoencoders.

\subsection{AE Training}

We train Video DC-AE with two different channel sizes, namely, 256 channels and 128 channels from scratch. Following DC-AE, our training loss has no KL loss component. 
We first train the Video DC-AE with reconstruction loss $L_1$ and perceptual loss $L_\text{LPIPS}$ for 250k steps:
\[ 
    \mathcal{L} = \mathcal{L}_1 + 0.5 \mathcal{L}_\text{LPIPS},
\]
then add an adversarial loss component, $L_\text{adv}$, for another 200k steps:
\[
    \mathcal{L} = \mathcal{L}_1 + 0.5 \mathcal{L}_\text{LPIPS} + 0.05\mathcal{L}_\text{adv}.
\]

We train each AE model on 8 GPUs with a local batch size of 1. 
Our training data consists of 32 frames of 256px videos at an aspect ratio of 1:1. 
We use a fixed learning rate of 5e-5 for the AE using the AdamW optimizer with $\beta$ values of (0.9, 0.999) and an $\epsilon$ value of $1\times 10^{-15}$ without weight decay.
The learning rate is adjusted to 1e-4 for the discriminator without changing any other optimizer parameters.
We apply gradient norm clipping at a threshold of 1.

\section{Conditioning}
\label{sec:condition}

\subsection{Image-to-Video Training}

\begin{figure}[h]
    \begin{center}
        \centerline{\includegraphics[width=\columnwidth]{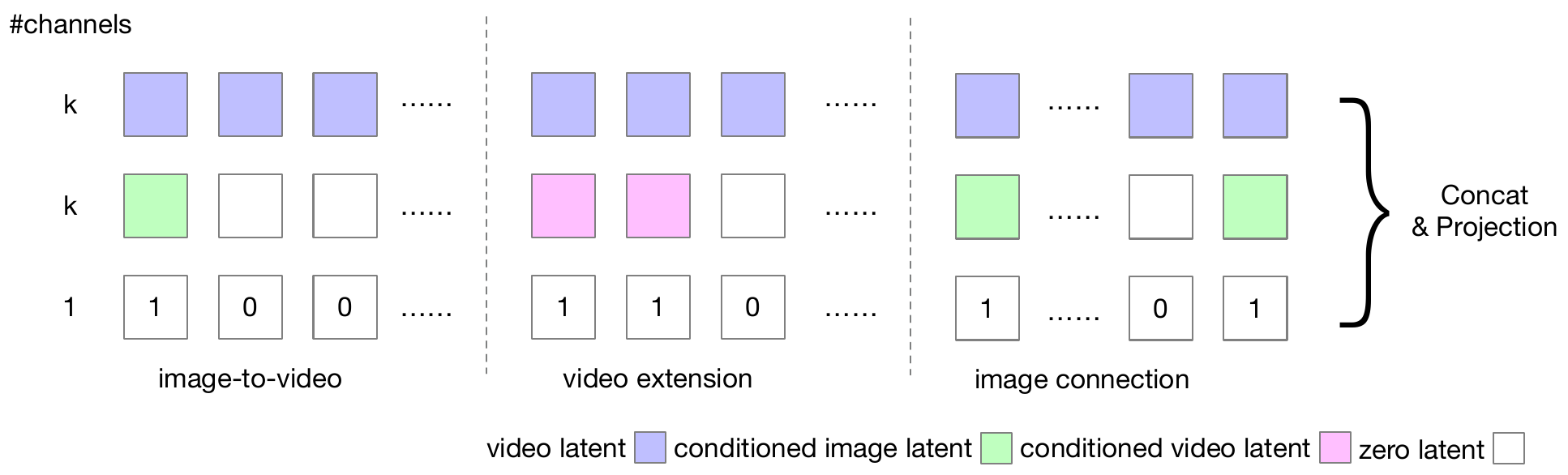}}
        \caption{Image and Video Condition Framework. Condition information is injected into the model via an additional channel dimension. A mask mechanism is introduced to distinguish different input types. This design enables support for a wide range of I2V and video-to-video (V2V) tasks.}
        \label{fig:i2v}
    \end{center}
    \vskip -20pt
\end{figure}

Open-Sora 1.2 introduces a universal conditioning framework, allowing both image and video conditions to be applied at any frame. However, its original method replaces noisy inputs with the condition, leading to inconsistent timesteps across different inputs. To address this issue, we modify the framework by concatenating the condition as additional channels, ensuring that the velocity prediction task remains unchanged.

As illustrated in Figure~\ref{fig:i2v}, our approach first encodes image or video conditions using an autoencoder, then concatenates the encoded features with the original video latent representation. An extra channel is introduced to indicate the task type, increasing the number of channels from  $k$ to $2k+1$ .

To improve generalization, we introduce image condition dropout similar to text condition dropout. During training, dropping the image condition reduces the problem to a T2V setting, where zero tensors are concatenated to the video latent. For T/I2V training, we set the dropout ratio to 12.5\%, ensuring robustness across various I2V and T2V tasks.

\subsection{Image-to-Video Inference}

\begin{figure}[t]
    \begin{center}
        \centerline{\includegraphics[width=0.75\columnwidth]{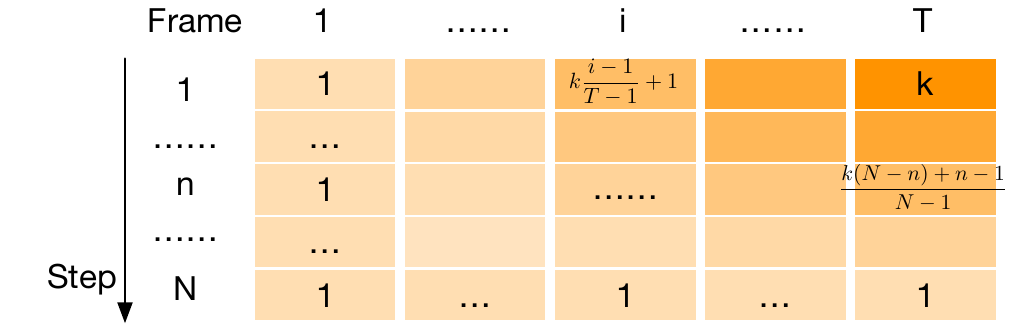}}
        \caption{Heatmap of Image Guidance Scale Across Denoising Steps and Latent Frames. Darker regions indicate higher image guidance values, emphasizing stronger influence on later frames and earlier denoising steps.}
        \label{fig:image-guidance}
    \end{center}
    \vskip -20pt
\end{figure}

\begin{figure}[t]
    \begin{center}
        \centerline{\includegraphics[width=0.9\columnwidth, trim={0cm 5cm 0cm 5cm}, clip]{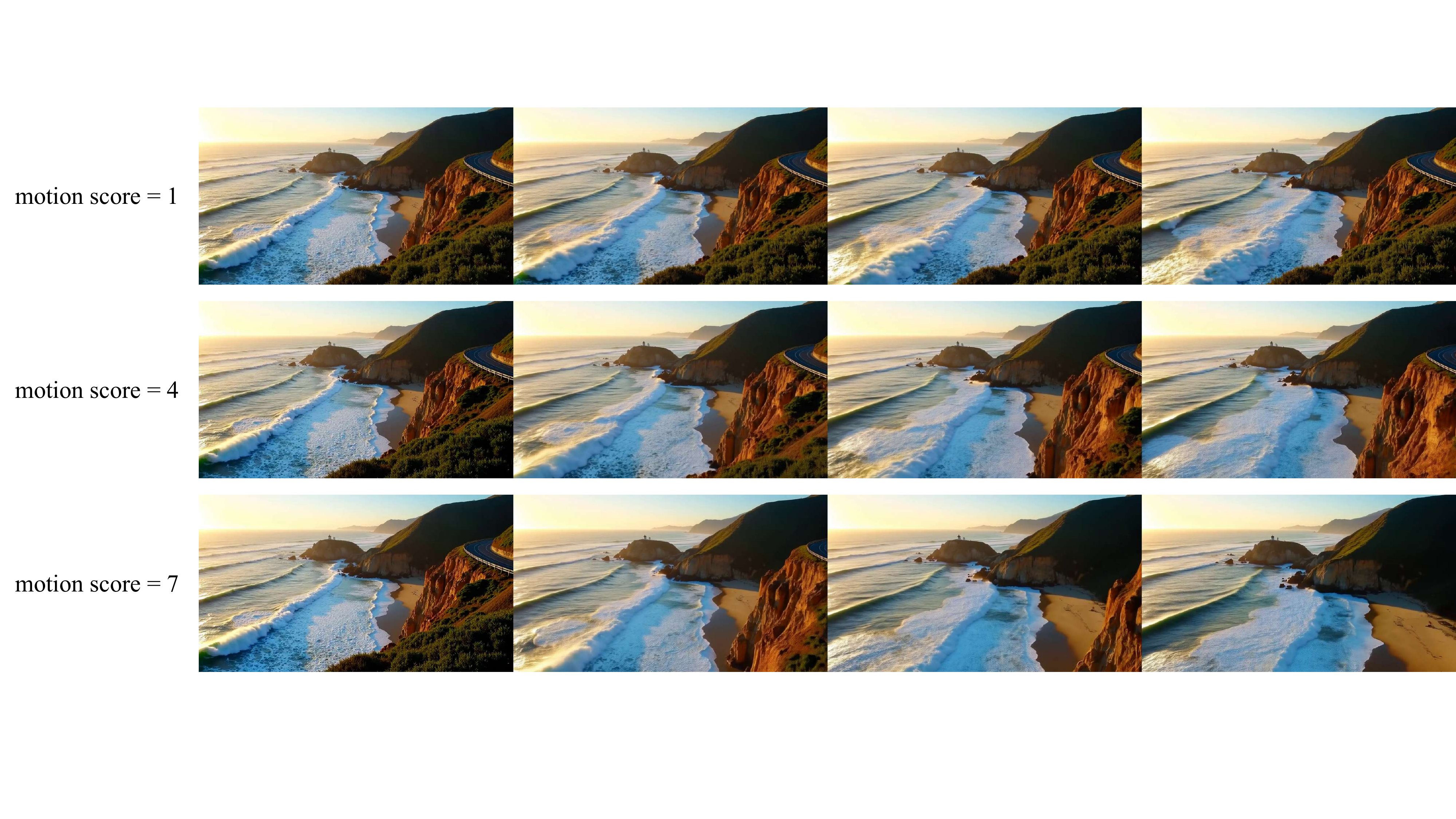}}
        \caption{Effect of Motion Score on Video Generation. This figure illustrates the impact of different motion scores on the generated video. As the motion score increases, the camera movement becomes more pronounced and the overall dynamic movement increases within the scene.}
        \label{fig:motion_score_performance}
    \end{center}
    \vskip -20pt
\end{figure}

We use classifier-free guidance for inference \cite{ho2022classifier}. As our model is conditioned on both image and text, a straightforward approach is to use a single guidance scale:
$$v_t = v_\theta(x_t,t,\varnothing,\varnothing) + g\cdot(v_\theta(x_t,t,\textit{txt},\textit{img})-v_\theta(x_t,t,\varnothing,\varnothing)),$$
where $v_\theta$ represents the predicted velocity, and $g$ is the guidance scale. However, we find that this approach is suboptimal. Image guidance requires only a small guidance scale, as large values make the entire video static, while text guidance benefits from a higher scale, improving semantic alignment. To address this, we decouple the guidance terms as follows:
\begin{align*}
    v_t &= v_\theta(x_t,t,\varnothing,\varnothing) \notag \\
    & + g_{\textit{img}}\cdot \big(v_\theta(x_t,t,\varnothing,\textit{img}) - v_\theta(x_t,t,\varnothing,\varnothing)\big) \notag \\
    &+ g_{\textit{txt}}\cdot \big(v_\theta(x_t,t,\textit{txt},\textit{img}) - v_\theta(x_t,t,\varnothing,\textit{img})\big).
\end{align*}

A further issue arises when using high image guidance, as it sometimes introduces flickering in generated videos. To mitigate this, we introduce a guidance oscillation technique~\cite{ho2022imagen} for image guidance. Taking 50-step sampling as an example, after the first 10 steps, we alternate the image guidance scale: during odd-numbered steps, we apply the original $g_{\textit{img}}$, while for even-numbered steps, we reduce it to 1. This helps balance stability and motion consistency.

Beyond oscillation, we introduce a dynamic image guidance scaling strategy. Since the image condition is primarily applied to the first frame, frames toward the end of the video require stronger image guidance to maintain coherence. At the same time, as denoising progresses, the video scene is mostly formed, making image guidance less critical at later diffusion steps. To optimize for both effects, we dynamically adjust $g_{\textit{img}}$  based on both the frame index and the denoising step. As shown in Figure~\ref{fig:image-guidance}, we test both linear and quadratic scaling, finding that linear scaling across both dimensions provides the best performance. In practice, we use default guidance values of $g_{\textit{img}}=3$ ,  $g_{\textit{txt}}=7.5$, achieving a balance between motion fidelity and semantic accuracy.

\subsection{Motion Score}
\label{sec:motion_score}
Controlling the level of motion strength is a crucial feature in video generation. 
% Depending on the scene, users may prefer high-fidelity videos with minimal motion or highly dynamic videos with significant movement. While techniques such as classifier-free guidance scaling and shifting denoising steps can influence motion intensity, we find that these approaches are often entangled with other factors, such as visual quality and text alignment, making precise control challenging. 
We explicitly model the motion dynamics as a separate controllable parameter. We leverage the motion score obtained from data pre-processing, which quantifies the level of dynamics in a video, by appending it to the caption as an additional conditioning signal. During inference, adjusting the motion score allows for effective and independent control over the dynamic level of the generated video, as shown in Figure~\ref{fig:motion_score_performance}.

\subsection{Inference-Time Scaling}
\label{sec_inference}

We additionally explore an inference-time scaling strategy inspired by \citet{ma2025inferencetimescalingdiffusionmodels} to enhance video generation quality and diversity without modifying the underlying model. Since this technique is not used in our main evaluation or comparative benchmarking, we defer the implementation details, and qualitative results to Appendix~\ref{sec:appendix_inference_scaling}.

\section{System Optimization}
\label{sec:system}

We train our models using ColossalAI~\cite{10.1145/3605573.3605613}, an efficient parallel training system. Our hardware setup includes H200 GPUs, whose 141GB memory enables more effective data parallelism (DP) and allows for more aggressive selective activation checkpointing, significantly optimizing resource utilization. Additionally, we leverage PyTorch compile \cite{Ansel_PyTorch_2_Faster_2024} and Triton kernels \cite{triton_language} to accelerate training efficiency.

To efficiently handle high-resolution video training, we employ multiple parallelization techniques. For video autoencoders, we adapt tensor parallelism (TP)~\cite{shoeybi2020megatronlmtrainingmultibillionparameter} to convolution layers by partitioning weights either the input or output channel dimensions, reducing memory consumption while preventing out-of-bound indexing for high-resolution training. For MMDiT training, we combine Zero Redundancy Optimizer (ZeroDP)~\cite{rajbhandari2020zeromemoryoptimizationstraining} with Context Parallelism (CP)~\cite{liu2023ringattentionblockwisetransformers}. Here, video and text sequences are partitioned across GPUs along the sequence dimension, enabling each GPU to compute attention independently. This approach mitigates the memory bottleneck and improves attention computation efficiency, particularly for high-resolution videos where attention complexity grows quadratically.

Empirical evaluations on H200 GPUs (141GB memory) indicate that employing CP alone yields an optimal trade-off between memory efficiency and computational performance. During Stage 1 and Stage 2 training, we exclusively utilize DP in combination with ZeRO-2, achieving a maximum FLOPs utilization (MFU) of 38.19\%. In Stage 3, we integrate ZeRO-2 with CP=4, resulting in a MFU of 35.75\%.

Beyond parallelization, our training system incorporates several established optimization techniques, including selective activation checkpointing, automatic failure recovery, using pinned-memory buffers for the dataloader, and the saving and loading of checkpoints. We defer detailed descriptions of these auxiliary optimizations to Appendix~\ref{sec:system_appendix}.

\section{Performance}
\label{sec:performance}

Our model supports both T2V and I2V generation at 256×256px and 768×768px resolutions (hereafter referred to as 256px and 768px), generating videos up to 128 frames long. At 24 FPS, this corresponds to a 5-second duration. Since our model is optimized for I2V generation, the default text-to-image-to-video (T2I2V) pipeline first generates an image using the FLUX model, which is then used as the starting frame for video generation. 
% Inference scaling is not used for evaluation.
The generated results are presented in Figure \ref{fig:video_display} in Appendix~\ref{sec:generation_samples}.

To benchmark our model against other approaches, we generate videos using Open-Sora 2.0, as well as several closed-source APIs and open-source models, using a set of 100 text prompts. To ensure fairness, we only perform a single inference per model, avoiding any cherry-picking. For all models, we use their default settings, with video lengths ranging from 5 to 6 seconds and resolutions between 768px and 720p, depending on their generation constraints.

A blinded evaluation was conducted by 10 professional evaluators, assessing videos based on three key criteria:
\begin{itemize}
    \item \textbf{Visual Quality}: Which video exhibits higher visual fidelity and is aesthetically more pleasing?
    \item \textbf{Prompt Adherence}: Which video aligns more accurately with the given text prompt?
    \item \textbf{Motion Quality}: Which video maintains more consistent motion and better adheres to physical laws?
\end{itemize}
The evaluation results, shown in Figure~\ref{fig:winrate}, indicate that our model outperforms existing models in several dimensions and achieves competitive performance across all categories.

\begin{figure}[h]
    \begin{center}
        \centerline{\includegraphics[width=\columnwidth]{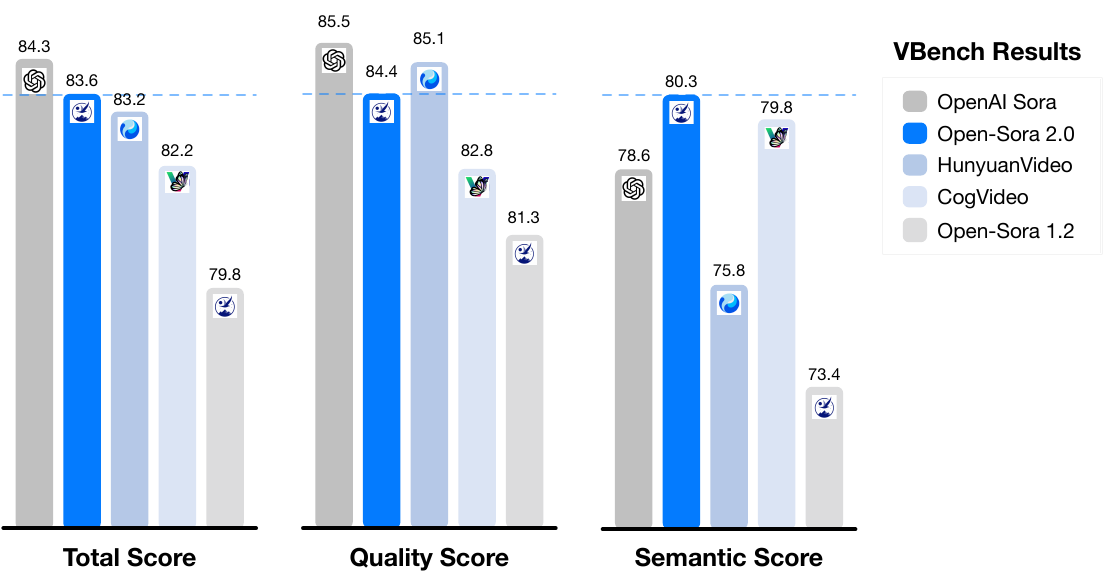}}
        \caption{VBench Score Comparison. Our model outperforms open-source T2V models by leveraging a text-to-image-to-video generation approach. Additionally, our latest version significantly narrows the performance gap between Open-Sora and OpenAI’s Sora, demonstrating substantial improvements in video generation quality and coherence.}
        \label{fig:vbench}
    \end{center}
    \vskip -20pt
\end{figure}

We further evaluate our model’s performance using VBench~\cite{huang2024vbench} in Figure~\ref{fig:vbench}, demonstrating significant improvements from Open-Sora 1.2 to 2.0. The performance gap between Open-Sora and OpenAI’s Sora has been reduced from 4.52\% to 0.69\%, highlighting substantial advancements in video generation quality. Additionally, our model achieves a higher VBench score compared to CogVideoX1.5-5B and HunyuanVideo, further establishing its superiority among current open-source T2V models.

% \begin{table}[h]
% \centering
% \resizebox{0.8\textwidth}{!}{%
% \begin{tabular}{@{}lccc@{}}
% \toprule
% Model           & Total Score & Quality Score & Semantic Score \\ \midrule
% Open-Sora 1.2  & 79.76\%     & 81.35\%       & 73.39\%        \\
% CogVideoX1.5-5B & 82.17\%     & 82.78\%       & \underline{79.76}\%        \\
% HunyuanVideo    & \underline{83.24}\%     & \underline{85.09}\%       & 75.82\%        \\ \midrule
% Open-Sora 2.0  & \textbf{83.59}\%     & 84.41\%       & \textbf{80.34}\%        \\
% Sora            & 84.28\%     & 85.51\%       & 78.57\%      
% \\ \bottomrule
% \end{tabular}%
% }
% \vspace{5pt}
% \caption{\textbf{VBench Score Comparison.} Our model outperforms open-source text-to-video models by leveraging a text-to-image-to-video generation approach. Additionally, our latest version significantly narrows the performance gap between Open-Sora and OpenAI’s Sora, demonstrating substantial improvements in video generation quality and coherence.}
% \label{tab:vbench}
% \end{table}
\section{Conclusion}

This paper introduces Open-Sora 2.0, a commercial-level video generation model that was trained for only \$200k, which is 5-10 times more cost-efficient than comparable models like MovieGen and Step-Video-T2V. This achievement highlights that high-quality video generation models can be developed with highly controlled costs through careful optimization of data curation, model architecture, training strategy, and system optimization. Despite its significantly lower training cost, Open-Sora 2.0 performs comparably to leading video generation models including HunyuanVideo and Runway Gen-3 Alpha. The model supports both T2V and I2V generation at resolutions up to 768×768 pixels for videos up to 5 seconds in length.

Looking ahead, several challenges remain in video generation. First, deep compression video VAE technology is still underexplored. While we aggressively increase the compression ratio to reduce latent tokens, this approach introduces reconstruction quality loss and adaptation difficulties. Additionally, diffusion models often produce artifacts such as object distortion and unnatural physics, with users having limited control over these details. The field needs further research on artifact prevention and enhanced control over generated content. We hope that by open-sourcing Open-Sora 2.0, we can provide the community with tools to collectively tackle these challenges, fostering innovation and advancements in the field of video generation.

%==============================================================================
% IMPACT STATEMENT
%==============================================================================
\section{Impact Statement}

This paper presents work whose goal is to advance the field of machine learning. There are many potential societal consequences of our work, none of which we feel must be specifically highlighted here.

\bibliography{ref}

@misc{ho2022imagen,
    title={Imagen Video: High Definition Video Generation with Diffusion Models},
    author={Jonathan Ho and William Chan and Chitwan Saharia and Jay Whang and Ruiqi Gao and Alexey Gritsenko and Diederik P. Kingma and Ben Poole and Mohammad Norouzi and David J. Fleet and Tim Salimans},
    year={2022},
    eprint={2210.02303},
    archivePrefix={arXiv},
    primaryClass={cs.CV}
}

@article{hoffmann2022training,
  title={Training compute-optimal large language models},
  author={Hoffmann, Jordan and Borgeaud, Sebastian and Mensch, Arthur and Buchatskaya, Elena and Cai, Trevor and Rutherford, Eliza and Casas, Diego de Las and Hendricks, Lisa Anne and Welbl, Johannes and Clark, Aidan and others},
  journal={arXiv preprint arXiv:2203.15556},
  year={2022}
}

@article{kaplan2020scaling,
  title={Scaling laws for neural language models},
  author={Kaplan, Jared and McCandlish, Sam and Henighan, Tom and Brown, Tom B and Chess, Benjamin and Child, Rewon and Gray, Scott and Radford, Alec and Wu, Jeffrey and Amodei, Dario},
  journal={arXiv preprint arXiv:2001.08361},
  year={2020}
}

@article{polyak2024moviegen,
  title={Movie Gen: A Cast of Media Foundation Models},
  author={Polyak, Adam and Zohar, Amit and Brown, Andrew and Tjandra, Andros and Sinha, Animesh and Lee, Ann and Vyas, Apoorv and Shi, Bowen and Ma, Chih-Yao and Chuang, Ching-Yao and Yan, David and Choudhary, Dhruv and Wang, Dingkang and Sethi, Geet and Pang, Guan and Ma, Haoyu and Misra, Ishan and Hou, Ji and Wang, Jialiang and Jagadeesh, Kiran and Li, Kunpeng and Zhang, Luxin and Singh, Mannat and Williamson, Mary and Le, Matt and Yu, Matthew and Singh, Mitesh Kumar and Zhang, Peizhao and Vajda, Peter and Duval, Quentin and Girdhar, Rohit and Sumbaly, Roshan and Rambhatla, Sai Saketh and Tsai, Sam and Azadi, Samaneh and Datta, Samyak and Chen, Sanyuan and Bell, Sean and Ramaswamy, Sharadh and Sheynin, Shelly and Bhattacharya, Siddharth and Motwani, Simran and Xu, Tao and Li, Tianhe and Hou, Tingbo and Hsu, Wei-Ning and Yin, Xi and Dai, Xiaoliang and Taigman, Yaniv and Luo, Yaqiao and Liu, Yen-Cheng and Wu, Yi-Chiao and Zhao, Yue and Kirstain, Yuval and He, Zecheng and He, Zijian and Pumarola, Albert and Thabet, Ali and Sanakoyeu, Artsiom and Mallya, Arun and Guo, Baishan and Araya, Boris and Kerr, Breena and Wood, Carleigh and Liu, Ce and Peng, Cen and Vengertsev, Dimitry and Schonfeld, Edgar and Blanchard, Elliot and Juefei-Xu, Felix and Nord, Fraylie and Liang, Jeff and Hoffman, John and Kohler, Jonas and Fire, Kaolin and Sivakumar, Karthik and Chen, Lawrence and Yu, Licheng and Gao, Luya and Georgopoulos, Markos and Moritz, Rashel and Sampson, Sara K. and Li, Shikai and Parmeggiani, Simone and Fine, Steve and Fowler, Tara and Petrovic, Vladan and Du, Yuming},
  journal={arXiv preprint arxiv:2410.13720},
  year={2024}
}

@inproceedings{cai2023efficientvit,
  title={Efficientvit: Lightweight multi-scale attention for high-resolution dense prediction},
  author={Cai, Han and Li, Junyan and Hu, Muyan and Gan, Chuang and Han, Song},
  booktitle={Proceedings of the IEEE/CVF international conference on computer vision},
  pages={17302--17313},
  year={2023}
}

@article{ma2025step,
  title={Step-Video-T2V Technical Report: The Practice, Challenges, and Future of Video Foundation Model},
  author={Ma, Guoqing and Huang, Haoyang and Yan, Kun and Chen, Liangyu and Duan, Nan and Yin, Shengming and Wan, Changyi and Ming, Ranchen and Song, Xiaoniu and Chen, Xing and others},
  journal={arXiv preprint arXiv:2502.10248},
  year={2025}
}

@article{kong2024hunyuanvideo,
  title={Hunyuanvideo: A systematic framework for large video generative models},
  author={Kong, Weijie and Tian, Qi and Zhang, Zijian and Min, Rox and Dai, Zuozhuo and Zhou, Jin and Xiong, Jiangfeng and Li, Xin and Wu, Bo and Zhang, Jianwei and others},
  journal={arXiv preprint arXiv:2412.03603},
  year={2024}
}

@article{chen2024deep,
  title={Deep compression autoencoder for efficient high-resolution diffusion models},
  author={Chen, Junyu and Cai, Han and Chen, Junsong and Xie, Enze and Yang, Shang and Tang, Haotian and Li, Muyang and Lu, Yao and Han, Song},
  journal={arXiv preprint arXiv:2410.10733},
  year={2024}
}

@article{yao2025reconstruction,
  title={Reconstruction vs. Generation: Taming Optimization Dilemma in Latent Diffusion Models},
  author={Yao, Jingfeng and Wang, Xinggang},
  journal={arXiv preprint arXiv:2501.01423},
  year={2025}
}

@article{xie2024sana,
  title={Sana: Efficient high-resolution image synthesis with linear diffusion transformers},
  author={Xie, Enze and Chen, Junsong and Chen, Junyu and Cai, Han and Tang, Haotian and Lin, Yujun and Zhang, Zhekai and Li, Muyang and Zhu, Ligeng and Lu, Yao and others},
  journal={arXiv preprint arXiv:2410.10629},
  year={2024}
}

@article{chen2025masked,
  title={Masked Autoencoders Are Effective Tokenizers for Diffusion Models},
  author={Chen, Hao and Han, Yujin and Chen, Fangyi and Li, Xiang and Wang, Yidong and Wang, Jindong and Wang, Ze and Liu, Zicheng and Zou, Difan and Raj, Bhiksha},
  journal={arXiv preprint arXiv:2502.03444},
  year={2025}
}

@article{videoworldsimulators2024,
  title={Video generation models as world simulators},
  author={Tim Brooks and Bill Peebles and Connor Holmes and Will DePue and Yufei Guo and Li Jing and David Schnurr and Joe Taylor and Troy Luhman and Eric Luhman and Clarence Ng and Ricky Wang and Aditya Ramesh},
  year={2024},
  url={https://openai.com/research/video-generation-models-as-world-simulators},
}

@article{ma2024latte,
  title={Latte: Latent diffusion transformer for video generation},
  author={Ma, Xin and Wang, Yaohui and Jia, Gengyun and Chen, Xinyuan and Liu, Ziwei and Li, Yuan-Fang and Chen, Cunjian and Qiao, Yu},
  journal={arXiv preprint arXiv:2401.03048},
  year={2024}
}

@article{lin2024open,
  title={Open-sora plan: Open-source large video generation model},
  author={Lin, Bin and Ge, Yunyang and Cheng, Xinhua and Li, Zongjian and Zhu, Bin and Wang, Shaodong and He, Xianyi and Ye, Yang and Yuan, Shenghai and Chen, Liuhan and others},
  journal={arXiv preprint arXiv:2412.00131},
  year={2024}
}

@article{chen2023pixart,
  title={Pixart-alpha: Fast training of diffusion transformer for photorealistic text-to-image synthesis},
  author={Chen, Junsong and Yu, Jincheng and Ge, Chongjian and Yao, Lewei and Xie, Enze and Wu, Yue and Wang, Zhongdao and Kwok, James and Luo, Ping and Lu, Huchuan and others},
  journal={arXiv preprint arXiv:2310.00426},
  year={2023}
}

@inproceedings{esser2024scaling,
  title={Scaling rectified flow transformers for high-resolution image synthesis},
  author={Esser, Patrick and Kulal, Sumith and Blattmann, Andreas and Entezari, Rahim and M{\"u}ller, Jonas and Saini, Harry and Levi, Yam and Lorenz, Dominik and Sauer, Axel and Boesel, Frederic and others},
  booktitle={Forty-first International Conference on Machine Learning},
  year={2024}
}

@article{lipman2022flow,
  title={Flow matching for generative modeling},
  author={Lipman, Yaron and Chen, Ricky TQ and Ben-Hamu, Heli and Nickel, Maximilian and Le, Matt},
  journal={arXiv preprint arXiv:2210.02747},
  year={2022}
}

@inproceedings{huang2024vbench,
  title={Vbench: Comprehensive benchmark suite for video generative models},
  author={Huang, Ziqi and He, Yinan and Yu, Jiashuo and Zhang, Fan and Si, Chenyang and Jiang, Yuming and Zhang, Yuanhan and Wu, Tianxing and Jin, Qingyang and Chanpaisit, Nattapol and others},
  booktitle={Proceedings of the IEEE/CVF Conference on Computer Vision and Pattern Recognition},
  pages={21807--21818},
  year={2024}
}

@article{hong2022cogvideo,
  title={Cogvideo: Large-scale pretraining for text-to-video generation via transformers},
  author={Hong, Wenyi and Ding, Ming and Zheng, Wendi and Liu, Xinghan and Tang, Jie},
  journal={arXiv preprint arXiv:2205.15868},
  year={2022}
}

@misc{shoeybi2020megatronlmtrainingmultibillionparameter,
      title={Megatron-LM: Training Multi-Billion Parameter Language Models Using Model Parallelism}, 
      author={Mohammad Shoeybi and Mostofa Patwary and Raul Puri and Patrick LeGresley and Jared Casper and Bryan Catanzaro},
      year={2020},
      eprint={1909.08053},
      archivePrefix={arXiv},
      primaryClass={cs.CL},
      url={https://arxiv.org/abs/1909.08053}, 
}

@misc{rajbhandari2020zeromemoryoptimizationstraining,
      title={ZeRO: Memory Optimizations Toward Training Trillion Parameter Models}, 
      author={Samyam Rajbhandari and Jeff Rasley and Olatunji Ruwase and Yuxiong He},
      year={2020},
      eprint={1910.02054},
      archivePrefix={arXiv},
      primaryClass={cs.LG},
      url={https://arxiv.org/abs/1910.02054}, 
}

@misc{liu2023ringattentionblockwisetransformers,
      title={Ring Attention with Blockwise Transformers for Near-Infinite Context}, 
      author={Hao Liu and Matei Zaharia and Pieter Abbeel},
      year={2023},
      eprint={2310.01889},
      archivePrefix={arXiv},
      primaryClass={cs.CL},
      url={https://arxiv.org/abs/2310.01889}, 
}

@software{FFmpeg,
  author = {{FFmpeg Developers}},
  title = {FFmpeg},
  url = {https://ffmpeg.org/},
  date = {2023},
  note = {Available from \url{https://ffmpeg.org/}}
}

@online{schuhmann2021aesthetic,
  author = {Schuhmann, Christoph},
  title = {Improved Aesthetic Predictor},
  year = {2021},
  url = {https://github.com/christophschuhmann/improved-aesthetic-predictor},
  urldate = {2025-03-10},
  organization = {GitHub}
}

@article{opencv_library,
    author = {Bradski, G.},
    title = {The OpenCV Library},
    journal = {Dr. Dobb's Journal of Software Tools},
    year = {2000}
}

@misc{paddleocr2020,
    title={PaddleOCR: A Practical Ultra Lightweight OCR System},
    author={PaddlePaddle Authors},
    howpublished = {\url{https://github.com/PaddlePaddle/PaddleOCR}},
    year={2020}
}

@software{pyscenedetect,
  author = {Castellano, Brandon},
  title = {PySceneDetect: Video Scene Cut Detection and Analysis Tool},
  url = {https://github.com/Breakthrough/PySceneDetect},
  version = {X.Y.Z},
  year = {2023}
}

@article{qwen25,
  title={Qwen2.5 technical report},
  author={Qwen Team},
  journal={arXiv preprint arXiv:2412.15115},
  year={2024}
}

@article{zhang2024video,
  title={Video instruction tuning with synthetic data},
  author={Zhang, Yuanhan and Wu, Jinming and Li, Wei and Li, Bo and Ma, Zejun and Liu, Ziwei and Li, Chunyuan},
  journal={arXiv preprint arXiv:2410.02713},
  year={2024}
}

@inproceedings{10.1145/3605573.3605613,
author = {Li, Shenggui and Liu, Hongxin and Bian, Zhengda and Fang, Jiarui and Huang, Haichen and Liu, Yuliang and Wang, Boxiang and You, Yang},
title = {Colossal-AI: A Unified Deep Learning System For Large-Scale Parallel Training},
year = {2023},
isbn = {9798400708435},
publisher = {Association for Computing Machinery},
address = {New York, NY, USA},
url = {https://doi.org/10.1145/3605573.3605613},
doi = {10.1145/3605573.3605613},
booktitle = {Proceedings of the 52nd International Conference on Parallel Processing},
pages = {766–775},
numpages = {10},
location = {Salt Lake City, UT, USA},
series = {ICPP '23}
}

@online{runwayml_gen3,
    author = {{RunwayML}},
    title = {Introducing {Gen-3 Alpha}},
    year = {2025},
    url = {https://runwayml.com/research/introducing-gen-3-alpha},
    organization = {RunwayML}
}

@online{luma_dream_machine,
    author = {{Luma AI}},
    title = {Dream Machine},
    year = {2025},
    url = {https://lumalabs.ai/dream-machine},
    organization = {Luma AI}
}

@article{zheng2024open,
  title={Open-sora: Democratizing efficient video production for all},
  author={Zheng, Zangwei and Peng, Xiangyu and Yang, Tianji and Shen, Chenhui and Li, Shenggui and Liu, Hongxin and Zhou, Yukun and Li, Tianyi and You, Yang},
  journal={arXiv preprint arXiv:2412.20404},
  year={2024}
}

@misc{ma2025inferencetimescalingdiffusionmodels,
      title={Inference-Time Scaling for Diffusion Models beyond Scaling Denoising Steps}, 
      author={Nanye Ma and Shangyuan Tong and Haolin Jia and Hexiang Hu and Yu-Chuan Su and Mingda Zhang and Xuan Yang and Yandong Li and Tommi Jaakkola and Xuhui Jia and Saining Xie},
      year={2025},
      eprint={2501.09732},
      archivePrefix={arXiv},
      primaryClass={cs.CV},
      url={https://arxiv.org/abs/2501.09732}, 
}

@misc{flux2024,
    author={Black Forest Labs},
    title={FLUX},
    year={2024},
    howpublished={\url{https://github.com/black-forest-labs/flux}},
}

@article{su2024roformer,
  title={Roformer: Enhanced transformer with rotary position embedding},
  author={Su, Jianlin and Ahmed, Murtadha and Lu, Yu and Pan, Shengfeng and Bo, Wen and Liu, Yunfeng},
  journal={Neurocomputing},
  volume={568},
  pages={127063},
  year={2024},
  publisher={Elsevier}
}

@article{chung2024scaling,
  title={Scaling instruction-finetuned language models},
  author={Chung, Hyung Won and Hou, Le and Longpre, Shayne and Zoph, Barret and Tay, Yi and Fedus, William and Li, Yunxuan and Wang, Xuezhi and Dehghani, Mostafa and Brahma, Siddhartha and others},
  journal={Journal of Machine Learning Research},
  volume={25},
  number={70},
  pages={1--53},
  year={2024}
}

@inproceedings{Radford2021LearningTV,
  title={Learning Transferable Visual Models From Natural Language Supervision},
  author={Alec Radford and Jong Wook Kim and Chris Hallacy and A. Ramesh and Gabriel Goh and Sandhini Agarwal and Girish Sastry and Amanda Askell and Pamela Mishkin and Jack Clark and Gretchen Krueger and Ilya Sutskever},
  booktitle={ICML},
  year={2021}
}

@inproceedings{zhang2018unreasonable,
  title={The unreasonable effectiveness of deep features as a perceptual metric},
  author={Zhang, Richard and Isola, Phillip and Efros, Alexei A and Shechtman, Eli and Wang, Oliver},
  booktitle={Proceedings of the IEEE conference on computer vision and pattern recognition},
  pages={586--595},
  year={2018}
}

@article{oquab2023dinov2,
  title={Dinov2: Learning robust visual features without supervision},
  author={Oquab, Maxime and Darcet, Timoth{\'e}e and Moutakanni, Th{\'e}o and Vo, Huy and Szafraniec, Marc and Khalidov, Vasil and Fernandez, Pierre and Haziza, Daniel and Massa, Francisco and El-Nouby, Alaaeldin and others},
  journal={arXiv preprint arXiv:2304.07193},
  year={2023}
}

@article{loshchilov2017decoupled,
  title={Decoupled weight decay regularization},
  author={Loshchilov, Ilya and Hutter, Frank},
  journal={arXiv preprint arXiv:1711.05101},
  year={2017}
}

@article{ho2022classifier,
  title={Classifier-free diffusion guidance},
  author={Ho, Jonathan and Salimans, Tim},
  journal={arXiv preprint arXiv:2207.12598},
  year={2022}
}

@inproceedings{Ansel_PyTorch_2_Faster_2024,
author = {Ansel, Jason and Yang, Edward and He, Horace and Gimelshein, Natalia and Jain, Animesh and Voznesensky, Michael and Bao, Bin and Bell, Peter and Berard, David and Burovski, Evgeni and Chauhan, Geeta and Chourdia, Anjali and Constable, Will and Desmaison, Alban and DeVito, Zachary and Ellison, Elias and Feng, Will and Gong, Jiong and Gschwind, Michael and Hirsh, Brian and Huang, Sherlock and Kalambarkar, Kshiteej and Kirsch, Laurent and Lazos, Michael and Lezcano, Mario and Liang, Yanbo and Liang, Jason and Lu, Yinghai and Luk, CK and Maher, Bert and Pan, Yunjie and Puhrsch, Christian and Reso, Matthias and Saroufim, Mark and Siraichi, Marcos Yukio and Suk, Helen and Suo, Michael and Tillet, Phil and Wang, Eikan and Wang, Xiaodong and Wen, William and Zhang, Shunting and Zhao, Xu and Zhou, Keren and Zou, Richard and Mathews, Ajit and Chanan, Gregory and Wu, Peng and Chintala, Soumith},
booktitle = {29th ACM International Conference on Architectural Support for Programming Languages and Operating Systems, Volume 2 (ASPLOS '24)},
doi = {10.1145/3620665.3640366},
month = apr,
publisher = {ACM},
title = {{PyTorch 2: Faster Machine Learning Through Dynamic Python Bytecode Transformation and Graph Compilation}},
url = {https://pytorch.org/assets/pytorch2-2.pdf},
year = {2024}
}

@misc{triton_language,
  author       = {OpenAI},
  title        = {Triton: An open-source programming language for writing highly efficient GPU code},
  year         = 2019,
  url          = {https://github.com/triton-lang/triton},
  note         = {Accessed: 2025-03-12}
}
\bibliographystyle{icml2026}

\newpage
\appendix
\onecolumn

\section{Data Statistics}
\label{sec:data_stat}

We conduct a statistical analysis of some key attributes of the video data, including aesthetic scores, duration (seconds), aspect ratios (height/width), and the length of the caption. To further explore the distribution of text content, we visualize common words in video captions using a word cloud.

\begin{figure*}[h]
    \begin{center}
        \centerline{\includegraphics[width=1\textwidth]{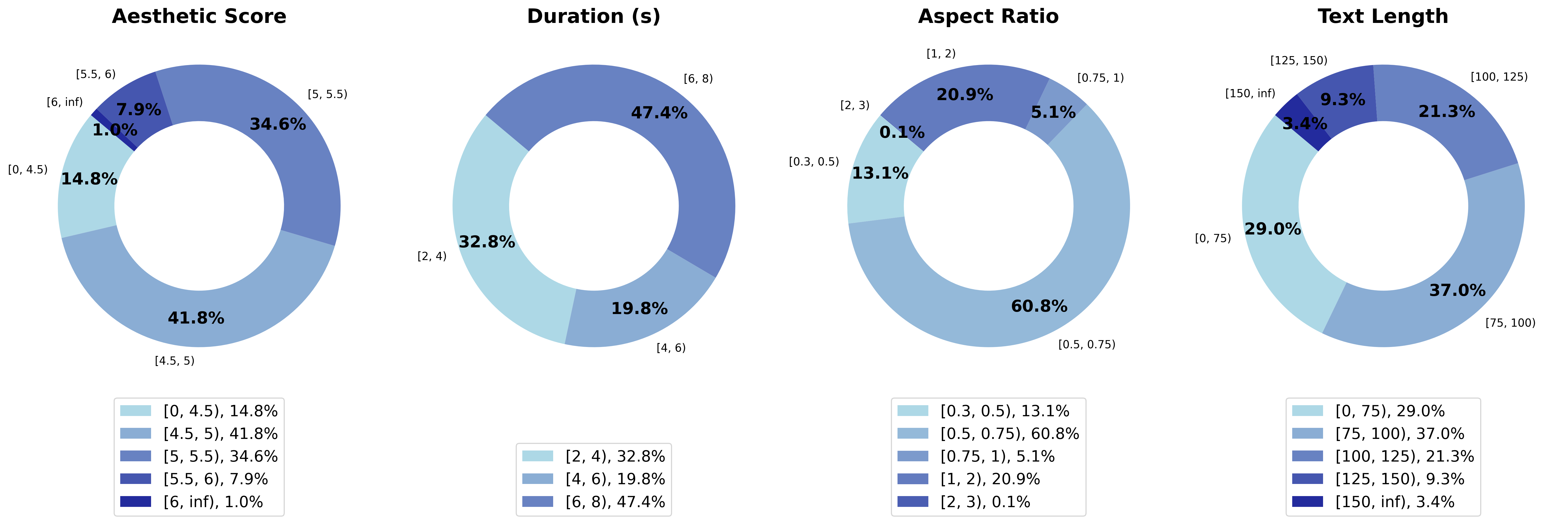}}
        \caption{Distribution of key attributes of the whole video dataset.}
        \label{fig:pie}
    \end{center}
\end{figure*}
\begin{figure}[h]
    \begin{center}
        \centerline{\includegraphics[width=0.7\columnwidth]{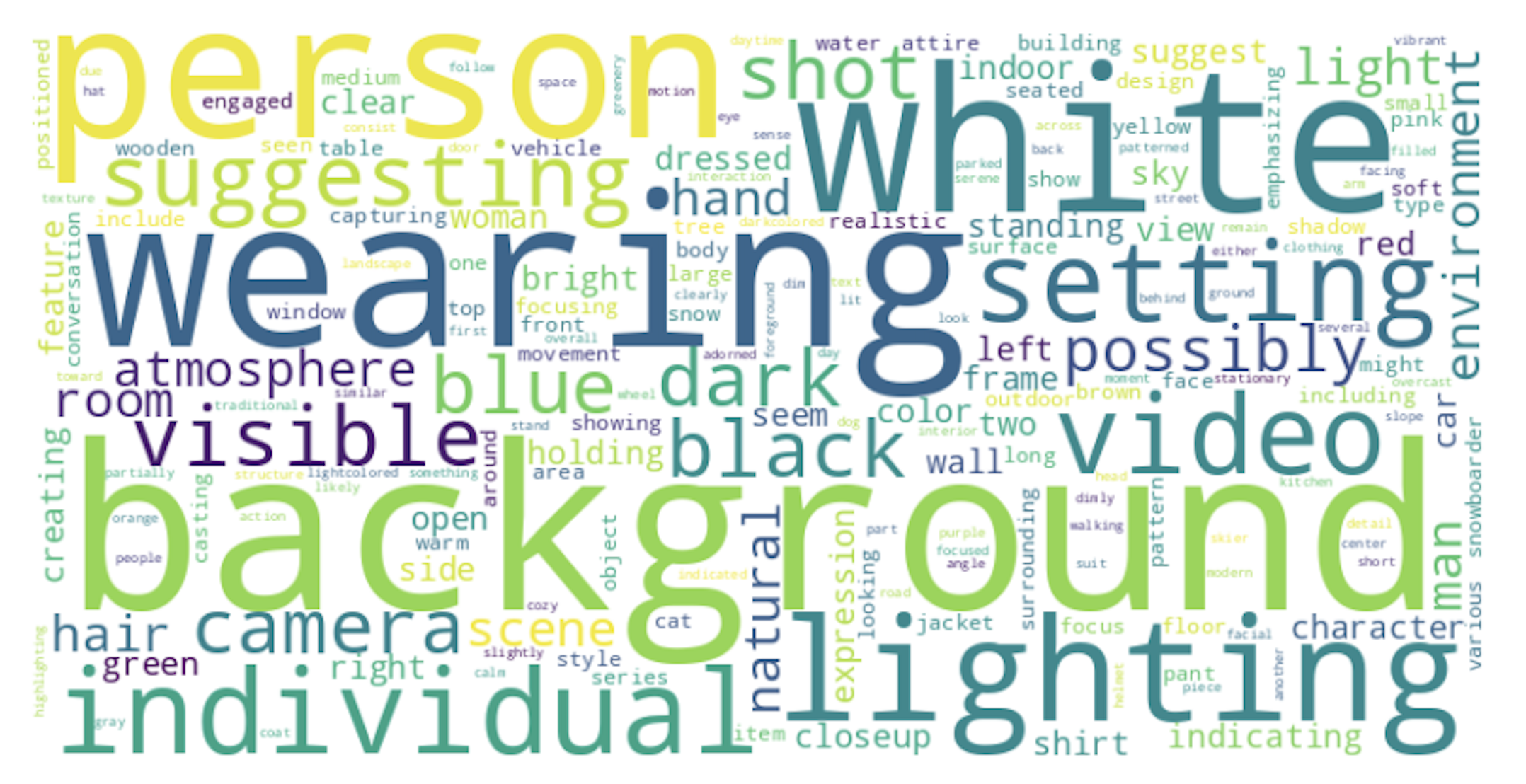}}
        \caption{Word cloud of the video captions.}
        \label{fig:caption_word_cloud}
    \end{center}
\end{figure}

As shown in Fig.~\ref{fig:pie}, the majority of videos have aesthetic scores ranging between 4.5 and 5.5, indicating a generally moderate level of visual appeal.
% The durations of the videos range from 2 to 8 seconds, with those between 6 and 8 seconds accounting for approximately half of the dataset.This preference for slightly longer clips provides richer temporal information for the model to learn dynamic patterns effectively.
Video durations range from 2 to 8 seconds, with nearly half of the dataset consisting of clips between 6 and 8 seconds, which offer richer temporal information for learning dynamic patterns.
% The statistics of video aspect ratios reveal that the majority fall between 0.5 and 0.75 (corresponding to 16:9). The diverse range of aspect ratios ensures that our model can adapt to different formats, improving its generalization ability. Over 70\% of video captions contain more than 75 words, providing detailed descriptions of the video content.
The analysis of aspect ratios shows that the majority fall between 0.5 and 0.75, corresponding to the 16:9 format, ensuring adaptability across different formats and enhancing the model’s generalization. Additionally, over 70\% of video captions exceed 75 words, providing detailed descriptions that contribute to a more informative training process.

% In Figure~\ref{fig:caption_word_cloud}, we demonstrate a word cloud of the vocabulary distribution in video captions. We find that captions not only include main subjects ("person") and their actions ("wearing"), but also background elements ("background", "setting", "atmosphere") and lighting conditions ("lighting"). This is in accordance with the 6 key aspects of our VLM prompting strategy introduced in Sec.~\ref{sec:data_ann}. Additionally, the word cloud suggests that a significant portion of the videos feature people, as indicated by the frequent appearance of "person" or "individual".
Figure~\ref{fig:caption_word_cloud} presents a word cloud illustrating the vocabulary distribution in video captions. The captions encompass not only main subjects such as \textit{“person”} and their actions like \textit{“wearing”}, but also background elements (\textit{“background”}, \textit{“setting”}, \textit{“atmosphere”}) and lighting conditions (\textit{“lighting”}). This aligns with the six key aspects of our VLM prompting strategy introduced in Sec.~\ref{sec:data_ann}. Furthermore, the frequent occurrence of \textit{“person”} and \textit{“individual”} suggests that a substantial portion of the dataset features human subjects.

\section{Data Filters}
\label{sec:data_filters}
\subsection{Aesthetic Score.} We assess the aesthetic quality of a sample using the CLIP+MLP aesthetic score predictor~\cite{schuhmann2021aesthetic}. For a video sample, we extract the first, middle, and last frame, compute their individual scores, and take the average as the final aesthetic score.

\subsection{Motion Score} 
% We evaluate the motion of a video using the VMAF motion score from FFmpeg's libavfilter library~\cite{FFmpeg}. Videos with either too low or too high motion scores are filtered.
The motion intensity of a video is measured using the VMAF motion score from FFmpeg’s libavfilter library~\cite{FFmpeg}. Videos with extremely low or excessively high motion scores are filtered out to maintain a balanced motion range.

\subsection{Blur Detection} 
% We evaluate the clarity of a sample by leveraging the Laplacian operator from OpenCV~\cite{opencv_library}. A sample is considered to be a blurry one if the variance of the Laplacian image is lower than a configurable threshold. For a video sample, 5 frames are extracted uniformly and the voting result is adopted.
To evaluate image clarity, we apply the Laplacian operator from OpenCV~\cite{opencv_library}. A sample is considered blurry if the variance of the Laplacian image falls below a configurable threshold. For video samples, we extract five uniformly spaced frames and adopt a majority voting approach to determine blurriness.

\subsection{OCR}
% We first detect text bounding boxes using PaddleOCR~\cite{paddleocr2020}. Then we calculate the total area of boxes of confidence higher than 0.7 as the final score. Images or videos with large text area are abandoned.
We detect text bounding boxes using PaddleOCR~\cite{paddleocr2020} and compute the total area of bounding boxes with a confidence score above 0.7. Images or videos containing excessive text are discarded to avoid unwanted overlays.

\subsection{Camera Jitter Detection}
% We use Shot Boundary Detection from the PySceneDetect library~\cite{pyscenedetect} to evaluate camera jitter. We recognize camera jitter as the average changing across frames is larger than a threshold. 
We detect camera jitter using Shot Boundary Detection from the PySceneDetect library~\cite{pyscenedetect}. A video is identified as having camera jitter if the average frame-to-frame change exceeds a predefined threshold.

\section{Model Architecture}
\subsection{Architecture Hyperparameters}
\label{sec:model_arch_hyperparameter}
\begin{table*}[h]
\centering
\resizebox{\textwidth}{!}{%
\begin{tabular}{@{}cccccc@{}}
\toprule
Double-Stream Layers & Single-Stream Layers & Model Dimension & FFN Dimension & Attention Heads & Patch Size \\ \midrule
19                                                & 38                                                & 3072            & 12288         & 24              & 2  \\ \bottomrule    
\end{tabular}%
}
\vspace{5pt}
\caption{Architecture hyperparameters for Open-Sora 2.0 11B parameter video generation model.}
\label{tab:model-arch}
\end{table*}

\section{Multi-Bucket Training}
\label{sec:multi-bucket-training}

For multi-bucket training, we adapt different batch sizes for different training configurations, such as parallelism strategy, video resolution, number of frames, and maximum number of frames. The specific batch sizes used for different training configurations are detailed in Table~\ref{tab:stage1-bs} and Table~\ref{tab:stage2-bs}.

\begin{table}[h]
\caption{Batch size and throughput in stage 1 and 2.}
\label{tab:stage1-bs}
  \begin{center}
    \begin{small}
      \begin{sc}
        \begin{tabular}{@{}ccccr@{}}
        \toprule
        Resolution             & \makecell{\# Frames}   & Max \# of Frames & Batch Size & Throughput on 8 GPUs \\ \midrule
        \multirow{4}{*}{256px} & 5 to 33    & 2304         & 12         & 12.7 videos/s     \\
                               & 37 to 65   & 4352         & 6          & 6.3 videos/s      \\
                               & 69 to 97   & 6400         & 4          & 4.2 videos/s      \\
                               & 101 to 129 & 8448         & 3          & 3.2 videos/s      \\ \midrule
        256px                  & 1          & 256          & 45         & 47.6 images/s     \\
        768px                  & 1          & 2304         & 13         & 13.8 images/s     \\
        1024px                 & 1          & 4096         & 7          & 7.4 images/s      \\ \bottomrule
        \end{tabular}%
    \end{sc}
    \end{small}
\end{center}
\vspace{10pt}
\end{table}

\begin{table}[h]
\caption{Batch size and throughput in stage 3 with context parallelism 4.}
\label{tab:stage2-bs}
\centering
  \begin{center}
    \begin{small}
      \begin{sc}
        \begin{tabular}{lcccr}
        \toprule
        Resolution             & \makecell{\# Frames}   & Max \# of Frames & Batch Size & Throughput on 8 GPUs \\ \midrule
        \multirow{4}{*}{768px} & 5 to 33    & 20736        & 6         & 0.25 videos/s    \\
                               & 37 to 65   & 39168         & 4          & 0.17 videos/s     \\
                               & 69 to 97   & 57600         & 3          & 0.13 videos/s     \\
                               & 101 to 129 & 76032         & 2          & 0.08  videos/s    \\ \midrule
        768px                  & 1          & 2304         & 38         & 1.60 images/s    \\ \bottomrule
        \end{tabular}%
    \end{sc}
    \end{small}
    \end{center}
\vspace{5pt}
\end{table}

To determine the batch sizes, we conduct a search on H200 GPUs. The process is as follows:

% Following Open-Sora 1.2, we adopt multi-bucket training. Specifically, we assign a batch size for videos with different numbers of frames and resolution. We sample videos at frame rates ranging from 16 to 24 FPS and incorporate the FPS information into the text prompt to provide an additional conditioning signal for video generation. Table~\ref{tab:stage1-bs} and table~\ref{tab:stage2-bs} shows the batch size for different training shapes. In this case, we can train on different lengths, resolution and aspect ratio of videos at the same time and maximize the GPU usage.

\begin{enumerate}
    \item We first identify the maximum batch size for the configuration with the highest token count, ensuring that it does not cause out-of-memory (OOM) errors.
    \item For all other configurations, batch sizes are chosen such that they are the largest possible without exceeding memory constraints, while ensuring that training time does not surpass that of the highest-token configuration.
    \item Additionally, we enforce the constraint that the combined execution time for autoencoder encoding and forward pass, as well as the backward pass, does not exceed the reference batch size’s execution time. This constraint helps prevent inefficiencies due to synchronous operations in the backward pass.
\end{enumerate}

By employing this strategy, we ensure efficient and scalable training across diverse video data distributions while maximizing hardware efficiency.

\section{Comparison of Video Generation with Different Autoencoder Compression Ratios}
\label{appendix:video_gen_ae}
\begin{figure*}[h]
  \vskip 0.2in
    \begin{center}
        \centerline{\includegraphics[width=0.9\textwidth]{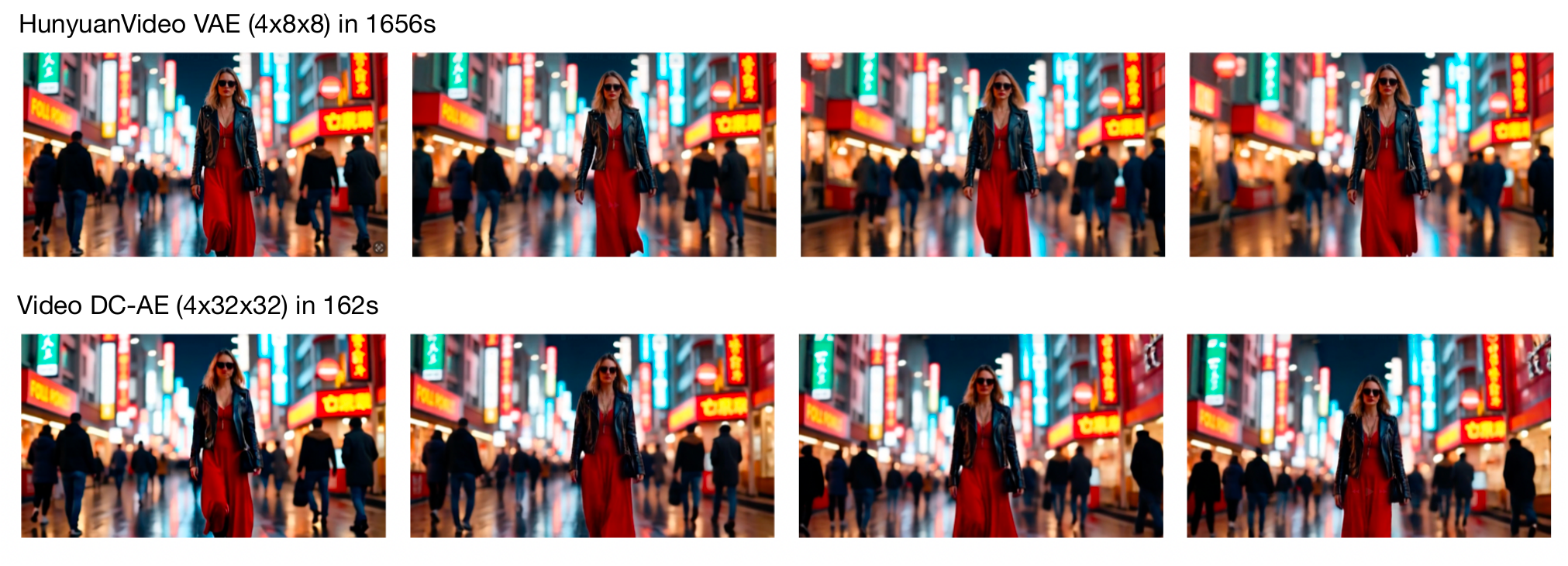}}
        \vspace{5pt}
        \caption{Comparison of Video Generation with Different Autoencoder Compression Ratios. The top and bottom rows correspond to lower and higher compression ratios, respectively. While the higher compression ratio AE results in slightly blurrier outputs, it significantly improves generation speed.}
        \label{fig:ae-comp}
    \end{center}
\end{figure*}

\section{High Compression Autoencoder Adaptation}
\subsection{Derivation of The Token Count Reduction Effects}
\label{sec:token_count_reduction}
Utilizing a high-compression video autoencoder significantly reduces the number of tokens required for video generation while increasing the number of latent channels incurs minimal additional computational cost (affecting only the input and output layers). 
We define the token downsample ratio, $D_{\text{token}}$, as the product of the autoencoder's compression ratios ($D_{\text{T | H | W}}$) and the generation model's patch sizes ($P_{\text{T | H | W}}$) in the T, H, W dimensions:
\[ 
    D_{\text{token}} = D_{\text{T}} \times D_{\text{H}} \times D_{\text{W}} \times P_{\text{T}} \times P_{\text{H}} \times P_{\text{W}}
\]
For instance, in a 5-second, 24fps, 768px video, the HunyuanVideo VAE with a $4\times 8 \times 8$ compression in combination of generation model's patch size $1 \times 2 \times 2$  ($D_{\text{token}} = 1024$) reduces the token count to 76K, whereas the Video DC-AE with $4\times 32 \times 32$  compression and patch size of $1 \times 1 \times 1$ has a $D_{\text{token}}$ of 4096,  effectively reduces the token count to 19K.

\section{Inference-Time Scaling}  

\label{sec:appendix_inference_scaling}

\begin{figure}[h]
    \centering
        \centerline{\includegraphics[width=0.7\columnwidth]{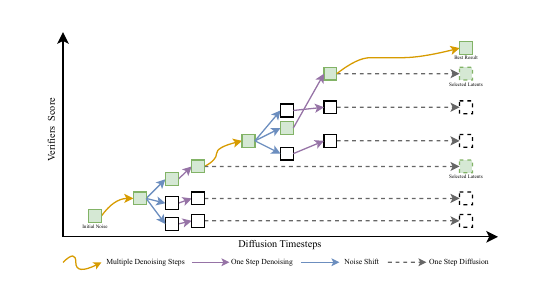}}
        \vspace{3pt}
    \caption{Diffusion Inference Scaling Framework. At selected denoising steps, controlled noise injections produce multiple candidate outputs. Each candidate is evaluated using VBench metrics, and the highest-scoring variant is chosen to continue the generation process. This enables quality-aware inference without modifying the model itself.}
    \label{fig:inference_scaling}
\end{figure}

We explore an inference-time scaling strategy inspired by \citet{ma2025inferencetimescalingdiffusionmodels} to enhance video generation quality and diversity without modifying the underlying model. As illustrated in Figure~\ref{fig:inference_scaling}, this approach introduces controlled noise variations during key denoising steps—particularly early steps (e.g., steps 1 and 3), which significantly influence the final output. At each selected step, partial denoising is performed to generate multiple candidate outputs, each corresponding to a different noise variant.

These candidates are then evaluated using six VBench metrics~\cite{huang2024vbench}:
\textit{subject\_consistency, background\_consistency, motion\_smoothness, dynamic\_degree, aesthetic\_quality, imaging\_quality}. Depending on the application, users can assign higher priority to specific metrics—for example, favoring \textit{motion\_smoothness} for more natural movements or \textit{aesthetic\_quality} for improved visual appeal. The highest-scoring candidate is selected to continue in the denoising process, allowing the model to explore a richer generation space with stronger alignment to target attributes.

The computational overhead of this strategy depends on four primary factors: the frequency of noise injection, the number of initial noise seeds, the number of variations per step, and the number of active metric verifiers. In our experiments, we fix the verifier set to isolate the effects of the other three variables.

\begin{figure}[h]
  \vskip 0.2in
    \centering
    \centerline{\includegraphics[width=0.7\columnwidth]{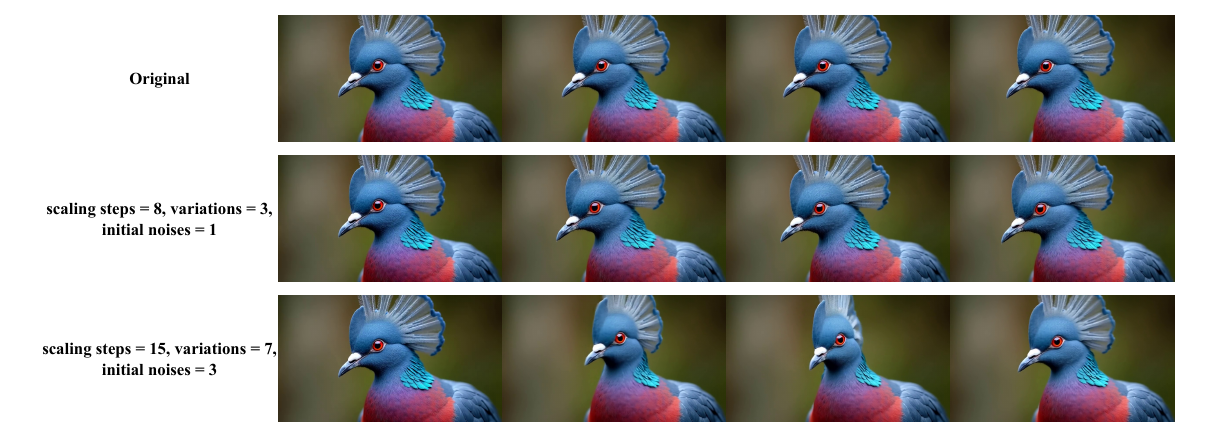}}
        \vspace{3pt}
    \caption{Inference Scaling Effects. Comparison of generated bird motion across three settings. \textbf{Top:} Baseline shows static or unnatural motion. \textbf{Middle:} Partial scaling provides limited enhancement. \textbf{Bottom:} Full scaling produces smooth and natural motion through quality-guided inference.}
    \label{fig:inference_scaling_effect}
\end{figure}

As shown in Figure~\ref{fig:inference_scaling_effect}, this method significantly improves generation quality under challenging prompts. The baseline (top row) exhibits static or unnatural motion, while moderate scaling (middle row) shows limited gains. In contrast, full scaling (bottom row) produces stable and natural motion sequences with improved visual coherence and dynamics.

\section{Other System Optimization}
\label{sec:system_appendix}
\subsection{Activation Checkpointing}

Activation checkpointing is applied selectively to reduce memory consumption without significantly increasing computational overhead. Instead of storing intermediate activations, only block inputs are retained, and the forward pass is recomputed during backpropagation. To minimize slowdowns, we avoid enabling checkpointing for every layer. In Stages 1 and 2, it is applied only to 8 layers of dual blocks and all single blocks, whereas in Stage 3 it is enabled for all blocks, supplemented by activation offloading to the CPU. The offloading mechanism further reduces memory footprint by asynchronously transferring activation tensors, leveraging pinned memory and asynchronous data movement to minimize training slowdowns.

\subsection{Auto Recovery}

To ensure continuous training in large-scale distributed environments, we implement an auto-recovery system to handle unexpected failures such as InfiniBand failures, storage system crashes, and NCCL errors. The system continuously monitors training status, checking for unresponsiveness, significant slowdowns, or loss spikes. Upon detecting issues, all processes are halted, and a faulty node checker diagnoses faulty nodes. If necessary, backup machines are deployed, and training resumes automatically from the last checkpoint without encountering a loss spike. With this strategy, our GPU utilization rate exceeds 99\%, minimizing downtime and optimizing hardware efficiency.

\subsection{Dataloader}

To accelerate data movement between the host (CPU) and devices (GPU), we optimize PyTorch’s dataloader. Instead of relying on PyTorch’s default pinned memory allocation, which invokes cudaMallocHost and may block CUDA kernel execution, we employ a pre-allocated pinned memory buffer to prevent dynamic memory allocations and reduce overhead, particularly for large video inputs. Furthermore, we overlap data transfers with computation, ensuring that data for the next step is prefetched while the current batch is being processed.

Additionally, we mitigate Python’s garbage collection (GC) overhead, which can unpredictably pause execution and cause severe inefficiencies in multi-process distributed training. To address this, we disable global GC and implement manual memory management, preventing unnecessary synchronization delays.

\subsection{Checkpoint Optimization}

Efficient model checkpointing is essential for minimizing recovery latency in distributed training. The checkpoint saving process involves three key steps: first, if the model is sharded, complete weights must be reconstructed via inter-GPU communication; second, the model weights are transferred from CUDA memory to CPU memory; and finally, the weights are written from the CPU to disk.

To optimize the second and third steps, we employ pre-allocated pinned memory to significantly accelerate weight transfer and implement asynchronous disk writing via C++, thereby ensuring that file I/O does not block the main training process. These optimizations reduce model-saving overhead to the order of seconds, which greatly accelerates the training.

For checkpoint loading, the process involves reading model weights from disk and transferring them from CPU to CUDA. We improve efficiency by using asynchronous pinned memory copying with multi-threaded allocation, and also implementing pipelined execution between shard reading and weight transfer. Furthermore, for large models stored in multiple shards, we overlap these phases across shards to maximize parallelism and accelerate the loading process. These optimizations reduce the overhead of training resumption.

\section{Generation Samples}
\label{sec:generation_samples}
\begin{figure}[h]
    \centering
    \begin{subfigure}[h]{1.0\columnwidth}
        \centering
        \includegraphics[width=0.4\columnwidth]{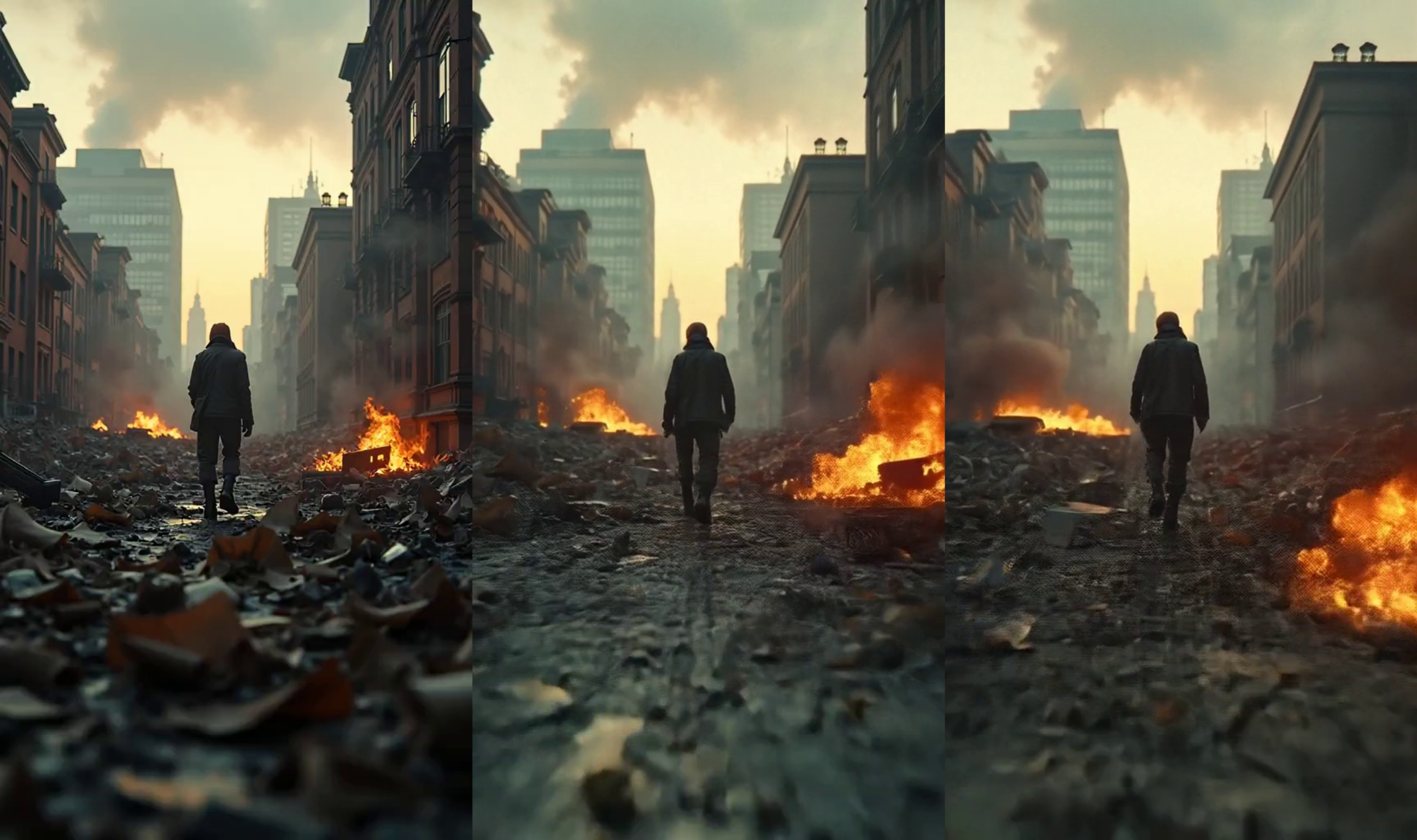}
        \caption{Prompt: A scene from a disaster movie.}
        \label{fig:ft_0055}
    \end{subfigure}
    
    \vspace{0.1cm} 
    
    \begin{subfigure}[h]{1.0\columnwidth}
        \centering
        \includegraphics[width=\columnwidth]{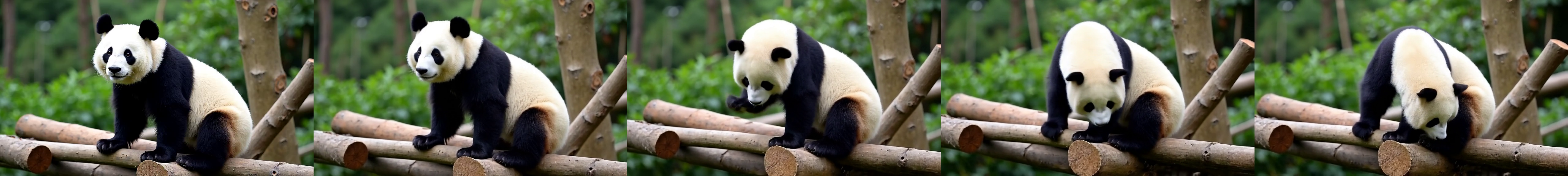}
        \caption{Prompt: A panda bear with distinct black patches climbs and rests on a wooden log platform amid lush, natural foliage.}
        \label{fig:douyin_0005}
    \end{subfigure}
    \vspace{0.1cm}
    \begin{subfigure}[h]{1.0\columnwidth}
        \centering
        \includegraphics[width=\columnwidth]{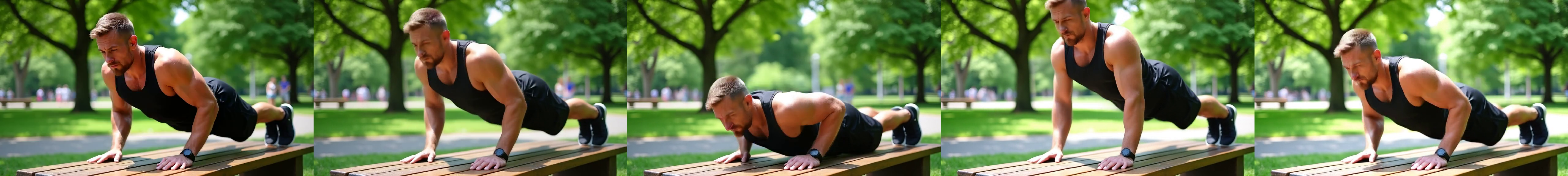}
        \caption{Prompt: A man performs push-ups on a wooden bench in a sunny park, captured from a side angle in a medium shot.
The focus is on his upper body and technique, with natural sunlight accentuating the scene.
Lush greenery and distant park-goers contribute to the energetic, realistic setting.}
        \label{fig:movie_0160}
    \end{subfigure}
    \begin{subfigure}[h]{1.0\columnwidth}
        \centering
        \includegraphics[width=\columnwidth]{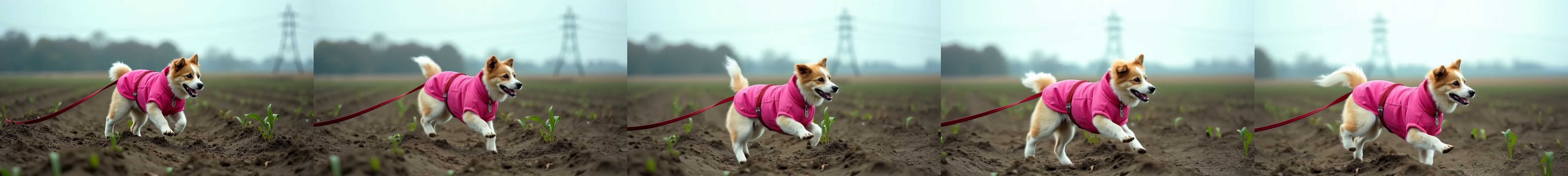}
        \caption{Prompt: A playful dog in a pink coat with a red leash dashes across a muddy field with sparse crops.
The camera tracks its energetic movement from right to left against a backdrop of trees and distant power lines under an overcast sky.
The realistic, medium shot captures a candid, lively moment in soft, diffused light.}
        \label{fig:movie_0463}
    \end{subfigure}    

    \vspace{0.1cm}

    \begin{subfigure}[h]{1.0\columnwidth}
        \centering
        \includegraphics[width=\columnwidth]{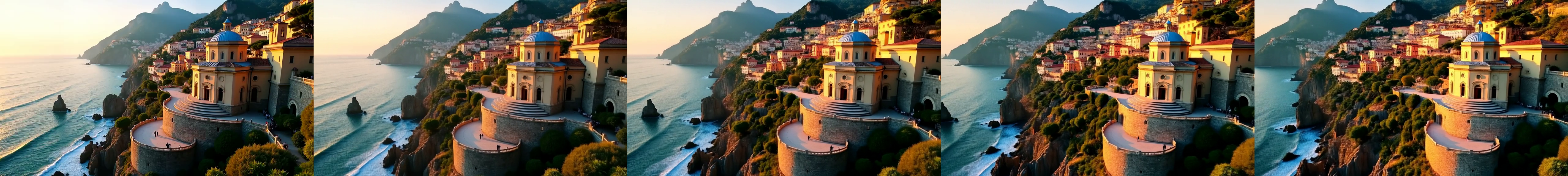}
        \caption{Prompt: A drone camera circles a historic church on a rocky outcrop along the Amalfi Coast, highlighting its stunning architecture, tiered patios, and the dramatic coastal views with waves crashing below and people enjoying the scene in the warm afternoon light.}
        \label{fig:sora_0019}
    \end{subfigure}    

    \vspace{0.1cm}

    \caption{High-quality videos generated by Open-Sora 2.0}
    \label{fig:video_display}
\end{figure}

\section*{Contributors}\label{team}

Core contributors are those who have been involved in the development of Open-Sora 2.0 throughout its entire process, while contributors are those who contributed part-time. All contributors are listed in \textbf{alphabetical order by first name}.

\begin{itemize}[left=0cm] 

\item \textbf{Project Leaders:} Xiangyu Peng, Zangwei Zheng.
\item \textbf{Core Contributors:}
    \begin{itemize}
        \item \textbf{Model \& Training:} Chenhui Shen, Tom Young, Xinying Guo.
        \item \textbf{Infrastructure:} Binluo Wang, Hang Xu, Hongxin Liu, Mingyan Jiang, Wenjun Li, Yuhui Wang.
        \item \textbf{Data \& Evaluation:} Anbang Ye, Gang Ren, Qianran Ma, Wanying Liang, Xiang Lian, Xiwen Wu, Yuting Zhong, Zhuangyan Li.
    \end{itemize}

\item \textbf{Contributors:} Chaoyu Gong, Guojun Lei, Leijun Cheng, Limin Zhang, Minghao Li, Ruijie Zhang, Silan Hu, Shijie Huang, Xiaokang Wang, Yuanheng Zhao, Yuqi Wang, Yuxuan Lou, Ziang Wei.

\item \textbf{Corresponding Authors:} Yang You (youy@comp.nus.edu.sg).

\end{itemize}
%%%%%%%%%%%%%%%%%%%%%%%%%%%%%%%%%%%%%%%%%%%%%%%%%%%%%%%%%%%%%%%%%%%%%%%%%%%%%%%
%%%%%%%%%%%%%%%%%%%%%%%%%%%%%%%%%%%%%%%%%%%%%%%%%%%%%%%%%%%%%%%%%%%%%%%%%%%%%%%

\end{document}